\documentclass[a4paper,12pt]{article}

\usepackage[T1]{fontenc}
\usepackage[latin2]{inputenc}
\usepackage{graphicx}
\input{epsfx}
\usepackage{subfig}
\usepackage{wrapfig}

\usepackage{amsmath}
\usepackage{amsfonts}
\usepackage{amssymb}
\usepackage{color}
\usepackage{rotating}

\newcommand{\be}{\begin{equation}}
\newcommand{\ee}{\end{equation}}
\newcommand{\bes}{\begin{subequations}}
\newcommand{\ees}{\end{subequations}}
\newcommand{\bea}{\begin{eqnarray}}
\newcommand{\eea}{\end{eqnarray}}
\newcommand{\bear}{\begin{equation}\begin{array}}
\newcommand{\eear}[1]{\end{array}\label{#1}\end{equation}}
\def\ba{$$\begin{array}}
\def\ea{\end{array}$$}
\def\bra{$\begin{array}}
 \def\era{\end{array}$}

\definecolor{Red}{rgb}{1,0,0}
\definecolor{Blue}{rgb}{0,0,1}

\newcommand{\fr}[2]{\dfrac{{ #1}}{{ #2}}}
\newsavebox{\fmbox}

{\end{list}}
\newcounter{enumct}

\begin{document}
\renewcommand{\tilde}{\widetilde}
\renewcommand{\ge}{\geqslant}
\renewcommand{\le}{\leqslant}

\title{Dark Matter data and constraints on quartic couplings in IDM}

\author{Dorota Soko\l owska \\
\textit{University of Warsaw, Faculty of Physics, Warsaw, Poland}
 }

\maketitle

\begin{abstract}
We analyse the thermal evolution of the Universe in the Inert Doublet Model for three regions of Dark Matter masses: low mass $(4-8)$ GeV, medium mass $(30-80)$ GeV and high mass $(500-1000)$ GeV. Those three regions of DM mass exhibit different behaviour, both in the possible types of evolution and in the energy relic density values. In this analysis we use the masses of the scalar particles as the input parameters to constrain the two self-couplings between neutral scalars: $\lambda_{345},\lambda_2$. These couplings are used to construct the parameter space where different types of evolution may be presented. We discuss the influence of the scalar masses on the type of the evolution. We also discuss the importance of the $\lambda_2$ self-coupling. We argue, that the astrophysical data along with the positivity constraints simultaneously constrain both $\lambda_{345}$ and $\lambda_2$ self-couplings. 
\end{abstract}

\section{Introduction}

Inert Doublet Model (IDM) is one of the widely discussed extensions of the Standard Model (SM) that may provide the Dark Matter (DM) candidate \cite{Deshpande:1977rw,Barbieri:2006dq}. It is a $Z_2$ symmetric 2HDM with a suitable set of parameters. In this model two scalar doublets are introduced. $\Phi_S$ is the Higgs doublet responsible for the electroweak symmetry breaking and masses of fermions and gauge bosons as in the SM. This doublet provides the longitudinal degrees of freedom of the gauge bosons and the SM-like Higgs boson particle $h_S$. The other doublet, $\Phi_D$, does not receive  vacuum expectation value (v.e.v.) and does not couple to fermions. All components of this doublet are realized as the massive scalar $D$-particles: two charged $D^\pm$ and two neutral
 $D_H$ and $D_{A}$. The stability of the lightest of these particles, which is a candidate for the DM particle, originates from the discrete symmetry of $Z_2$ type, called below $D$-symmetry:
\begin{equation}
D: \quad \Phi_S \xrightarrow{D} \Phi_S,\quad
	\Phi_D \xrightarrow{D} -\Phi_D,\quad
	 \textrm{SM fields}    \xrightarrow{D}  \textrm{SM fields}.
	\label{dtransf}
\end{equation}

We assume that the current vacuum state of the Universe is described by IDM with the DM candidate $D_H$, so that:
\be
M_{D^\pm}, M_{D_A} > M_{D_H}.
\label{chargedheavy}
\ee

In this paper we discuss the evolution of the Universe after inflation, following the approach presented in \cite{Ginzburg:2007jn, Ginzburg:2009dp, Ivanov:2008er, Ginzburg:2010wa,nasza_druga}. 
In this approach the scalar potential changes due to the temperature corrections $\sim T^2$ and in the past, for $T \not = 0$, the vacuum structure could have different properties than the present  vacuum state at $T=0$. Sequences of the different vacua (represented on the phase diagrams by \textit{rays}) describe different types of evolution of the Universe.

We consider the possible rays that can be realized in three regions of the DM mass: low mass $(4-8)$ GeV, medium mass $(30-80)$ GeV and high mass $(500-1000)$ GeV. We use the estimation of the 
energy relic density of DM to show that for different mass regions different types of evolution are consistent with the WMAP constraints for the energy relic density of DM, $\Omega_{DM} h^2$.

In \cite{Ginzburg:2010wa,nasza_druga} we focused on $(\mu_1,\mu_2)$ phase diagram defined by the quadratic and two quartic parameters of the scalar potential $V_{12} = -\left[m_{11}^2 \Phi_S^\dagger\Phi_S + m_{22}^2 \Phi_D^\dagger\Phi_D \right]/2 + \left[\lambda_1 \left(\Phi_S^\dagger\Phi_S\right)^2 + \lambda_2 \left(\Phi_D^\dagger\Phi_D \right)^2\right]/2$, in a following way:
\begin{eqnarray}
 &\mu_1 = m_{11}^2/\sqrt{\lambda_1} , \quad  \mu_2= m_{22}^2/\sqrt{\lambda_2}.& 
\end{eqnarray}

This phase diagram allowed us to present the possible types of evolution in easy way in form of rays. Each of the considered evolutions can be realized only if certain conditions for $\mu_1,\mu_2$ are satisfied. In the numerical analysis in \cite{nasza_druga} we have fixed the scalar masses, while the values of the scalar self-couplings $\lambda_2$ and $\lambda_{345}$ were chosen to fulfil the constraints for each ray. In this work we focus on those couplings, mainly on the $\lambda_2$ parameter, which is a quartic coupling for the DM particles, usually neglected in analysis of $\Omega_{DM} h^2$. However, this parameter has an important impact on the evolution and the history of the Universe.

The content of this paper is as follows. 
In section 2 we present the main properties of IDM. In section 3 we discuss in details the DM relic density constrains in three regions of DM mass. In section 4 we introduce the $(\lambda_{345},\lambda_2)$ plane, which is useful in the discussion of the parameter regions for different rays, at the fixed scalar masses. Here we use the benchmark points to illustrate three regions of DM mass and present the connection between the type of evolution and the values of $\Omega_{DM} h^2$. Section 5 contains a discussion on influence of different mass parameters on the $(\lambda_{345},\lambda_2)$ regions along with the relic density values.

\section{Model properties}

\subsection{IDM}

\paragraph{Lagrangian}
We consider
an electroweak symmetry breaking (EWSB) via the Brout-Englert-Higgs-Kibble (BEHK) mechanism described by the Lagrangian
\begin{equation}
{ \cal L}={ \cal L}^{SM}_{ gf } +{ \cal L}_H + {\cal L}_Y(\psi_f,\Phi_S) \,, \quad { \cal L}_H=T-V\, .
\label{lagrbas}
\end{equation}
Here, ${\cal L}^{SM}_{gf}$ describes the SM
interaction of gauge bosons and fermions.

The Higgs scalar Lagrangian ${\cal L}_H$ describes the interaction of two scalar  doublets $\Phi_S$ and $\Phi_D$ with the standard kinetic term $T$
and the scalar potential $V$. The potential $V$, which can describe IDM, is invariant under $D$-symmetry:
\begin{eqnarray}
V = -\frac{1}{2}\left[m_{11}^2 \Phi_S^\dagger\Phi_S + m_{22}^2 \Phi_D^\dagger\Phi_D \right] + \frac{1}{2}\left[\lambda_1 \left(\Phi_S^\dagger\Phi_S\right)^2 + \lambda_2 \left(\Phi_D^\dagger\Phi_D \right)^2\right] \nonumber \\
+  \lambda_3 \left(\Phi_S^\dagger\Phi_S \right) \left(\Phi_D^\dagger\Phi_D\right) + \lambda_4 \left(\Phi_S^\dagger\Phi_D\right) \left(\Phi_D^\dagger\Phi_S\right) +\frac{1}{2}\lambda_5\left[\left(\Phi_S^\dagger\Phi_D\right)^2\!+\!\left(\Phi_D^\dagger\Phi_S\right)^2\right]. \label{pot}
\end{eqnarray}
All parameters of (\ref{pot}) are real and one can fix $\lambda_5 <0$ without a loss of generality.

In order to have a stable vacuum we impose \textit{positivity constraints} in the following form:
\begin{eqnarray}
& \lambda_1>0\,, \quad \lambda_2>0, \quad R + 1 >0, & \label{posit}\\[1mm]
& \lambda_{345}=\lambda_3+\lambda_4+\lambda_5,\quad R = \lambda_{345}/\sqrt{\lambda_1 \lambda_2}.&
\end{eqnarray}
They assure that the potential is bounded from below and the extremum with the lowest energy will be the vacuum (the global minimum of the potential).

$D$-symmetric potential has 7 independent real parameters: $m_{11}^2,\, m_{22}^2,\, \lambda_{1-5}$. After EWSB those parameters can be expressed by the non-zero v.e.v, four scalar masses and two self-couplings between scalars.

${\cal L}_Y$ describes the Yukawa interaction of fermions $\psi_f$
with only one scalar doublet $\Phi_S$. It has the same form as in the SM with the change $\Phi\to\Phi_S$ (Model I for Yukawa interaction). ${\cal L}_Y$ respects $D$-symmetry in any order of the perturbation theory.

\paragraph{Inert vacuum}
We assume that the vacuum state of the potential (\ref{pot}) is given by the inert state (denoted by $I_1$). In this case only $\Phi_S$ acquires the non-zero vacuum expectation value, which as it follows from extremum condition is equal to:
\begin{equation}
v^2 = m_{11}^2/\lambda_1.
\end{equation}
For $I_1$ to be the vacuum following conditions should be satisfied \cite{Ginzburg:2010wa}:
\begin{eqnarray}
& \mu_{1} >0 \textrm{ for any } R, \quad \mu_1 > \mu_2 \textrm{ for } R>1, \quad R \mu_1 > \mu_2 \textrm{ for } |R|<1. &
\end{eqnarray}

The inert vacuum state is invariant under the $D$-transformation just as the whole basic Lagrangian \eqref{lagrbas}. There exist four dark scalar particles $D_H,\,D_A,
D^\pm$, which are $D$-odd, and the Higgs particle $h_S$, which interacts
with the fermions and gauge bosons just as the Higgs boson in the SM. $h_S$ and the SM fields are $D$-even. In the inert vacuum the $D$-parity is conserved, and  due to this fact
the lightest $D$-odd particle is stable, being a good  DM candidate.

The masses of the physical fields $h_S,\,D_H,\,D_A$ and
$D^\pm$ can be used to express the parameters of $V$ after EWSB. Those relations are given by:
\bear{c}
M_{h_s}^2=\lambda_1v^2= m_{11}^2\,,\qquad M_{D^\pm}^2=\fr{\lambda_3 v^2-m_{22}^2}{2}\,,\\[3mm]
M_{D_A}^2=M_{D^\pm}^2+\fr{\lambda_4-\lambda_5}{2}v^2\,,\qquad M_{D_H}^2=
M_{D^\pm}^2+\fr{\lambda_4+\lambda_5}{2}v^2\,.
\eear{massesA}

Other crucial parameters after EWSB are the couplings between the scalars. To complete the set of parameters we use two of them, $\lambda_{345}$ and $\lambda_2$. The $\lambda_{345}$ is a coupling between SM-like Higgs $h_S$ and DM candidate $D_H$: $D_H D_H h_S$ and $D_H D_H h_S h_S$. The $\lambda_2$ coupling is proportional to a \emph{quartic} self-coupling among $D$-particles, e.g. $D_H D_H D_H D_H$. The self-coupling that governs  the charged scalars' interactions: $D^+ D^- h_S$ and $D^+ D^- h_S h_S$ is proportional to  $\lambda_3$.

\subsection{Collider constraints}

Various  theoretical and experimental constraints apply to the IDM (see e.g.~\cite{Cao:2007rm, Agrawal:2008xz, Gustafsson:2007pc,Dolle:2009fn, Dolle:2009ft, LopezHonorez:2006gr, Arina:2009um, Tytgat:2007cv, Honorez:2010re, Lundstrom:2008ai, Krawczyk:2009fb}). 
Below we list important existing constraints on couplings and masses both for the Higgs particle and the dark scalars.

\paragraph{Constraints on self-couplings} The positivity constraints are imposed directly on quartic parameters in the potential. If we want to assure the perturbativity of the theory, the self-couplings $\lambda's$ (i.e. $\lambda_2,\lambda_3,\lambda_{345})$ cannot be large. The bound (called perturbativity constraint) is set typically to
\begin{equation}
|\lambda|<4 \pi.
\end{equation}

\paragraph{Electroweak precision tests} EWPT constrain strongly physics beyond SM. In IDM they limit the allowed values of masses both for the Higgs particle $h_S$ and the dark scalars. For IDM both light and heavy Higgs particle is allowed by EWPT \cite{Barbieri:2006dq}. Constraints for the dark scalars can be conveniently expressed by the limits for the mass splittings:
\begin{equation}
\delta_A = M_{D_A} - M_{D_H}, \quad \delta_{\pm} = M_{D^\pm} - M_{D_H}.
\end{equation}
 In \cite{Dolle:2009fn} it was obtained that for a light Higgs boson, the allowed region corresponds to  $\delta_{\pm} \sim \delta_{A} $ with mass splittings that could be large.
 For heavy SM Higgs large $\delta_{\pm}$ is needed, while $\delta_{A} $ could be small. In this work we limit ourselves to the light SM-like Higgs boson $h_S$.

\paragraph{LEP II limits} As $D^\pm, D_A, D_H$ do not couple to fermions, the LEP limits based on Yukawa interaction for the standard 2HDM don't apply. However, the signatures are similar to neutralinos and charginos interactions in MSSM. The absence of a signal within searches for supersymmetric neutralinos at LEP II was interpreted within the IDM  in paper \cite{Lundstrom:2008ai}. This analysis excludes the following region of masses: $M_{D_H} < 80$ GeV, $M_{D_A} < 100$ GeV and $\delta_A > 8$ GeV. For $\delta_A <8$ GeV the LEP I limit $M_{D_H} + M_{D_A} > M_Z$ applies.


\section{DM relic density constraints}

In this analysis we assume that $D_H$ is a dominant component of the observed DM, with its energy relic density in the Universe estimated to \cite{PDG}:

\begin{equation}
\Omega_{DM}h^2=0.112 \pm 0.009.\label{DMdens}
\end{equation}

Various studies show \cite{Dolle:2009fn, Dolle:2009ft, LopezHonorez:2006gr, Arina:2009um, Tytgat:2007cv, Honorez:2010re} that for IDM in most regions of the parameter space $\Omega_{DM}h^2$ is too low to fulfil the astrophysical constraints. However, there are three allowed regions of $M_{D_H}$: (i) light DM particles with mass close to and below $10 \textrm{ GeV}$, (ii) medium DM mass of $40-80 \textrm{ GeV}$ and (iii) heavy DM of mass larger than $500 \textrm{ GeV}$. 

Those regions are further constrained by the value of $\lambda_{345}$, which governs the main decay channels (decay through Higgs exchange) for $M_{D_H} < M_W$. 
In general, for larger $|\lambda_{345}|$ the relic density decreases due to the enhanced $D_H D_H$ annihilation via $h_S$ into pair of fermions (typically $\bar{b}b$). 

On the contrary, the value of $\lambda_{2}$ does not influence the DM relic density directly. At the same time, this parameter is difficult to access at colliders. Therefore, this parameter is usually fixed to abitrary small value in the DM analysis of IDM \cite{Dolle:2009fn, Dolle:2009ft, LopezHonorez:2006gr, Arina:2009um, Tytgat:2007cv, Honorez:2010re}. However, as we argue in this paper, value of $\lambda_2$ limits the value of $\lambda_{345}$ by the positivity constraints. Its value is also limited by the constraints arising from the possible existence of extrema and vacua of different properties than $I_1$ during the evolution of the Universe. 
Therefore, in this analysis we will consider the constraints for $M_{D_H}$ as well as for both $\lambda_{345}$ and $\lambda_2$ arising from the astophysical data.

In this work we derive  the corresponding relic-density constraints for IDM, using micrOMEGAs \cite{Belanger:2010gh}, with the IDM implemented by us. In analysis we respect all other existing limits and confirm findings of \cite{Dolle:2009fn, Dolle:2009ft, LopezHonorez:2006gr, Arina:2009um, Tytgat:2007cv, Honorez:2010re}. 
The micrOMEGAs program neglects temperature dependence of physical parameters and a possibility of more than one phase transition. In paper \cite{Ginzburg:2010wa} we concluded that if in the past there were sequences of phase transitions, then the Universe entered the inert phase with DM candidate at lower temperatures that in the typical one-stage EWSB. In principle, this should be considered while solving the Boltzmann equations for DM relic density, especially while the final phase transition into $I_1$ takes place at the low temperatures during the freeze-out.
Also the latent heat of the 1st-order transition 
may give significant corrections in some regions of the allowed parameter space \cite{pracaTEMP}. 
In this sense, the energy relic density calculations in this paper should be considered only as a preliminary estimate.\footnote{More work is in progress.}
 
Below we discuss the general trends in the change of $\Omega_{DM} h^2$ value 
in three regions of DM masses. In all examples we consider the light SM-like Higgs with its mass fixed to $M_h = 120 \textrm{ GeV}$. We expect the $\Omega_{DM} h^2$ to be in the 3$\sigma$ WMAP range, namely:
\begin{equation}
0.085 < \Omega_{DM} h^2 < 0.139. \label{omega}
\end{equation}
Note, that the medium DM mass region is  strongly constrained by the existing collider data.


\subsection{Low DM mass region \label{low_DM_sec}}

In the low mass region $M_{D_H}$ is of the order $(4-8) \textrm{ GeV}$ and masses of $D_A$ and $D^\pm$ are almost degenerate  with $\delta_A \approx \delta_{H^\pm} \approx 100 \textrm { GeV}$ \cite{Dolle:2009fn}.

\begin{figure}[htb]
\vspace{-10pt}
  \centering
 \includegraphics[width=0.6\textwidth]{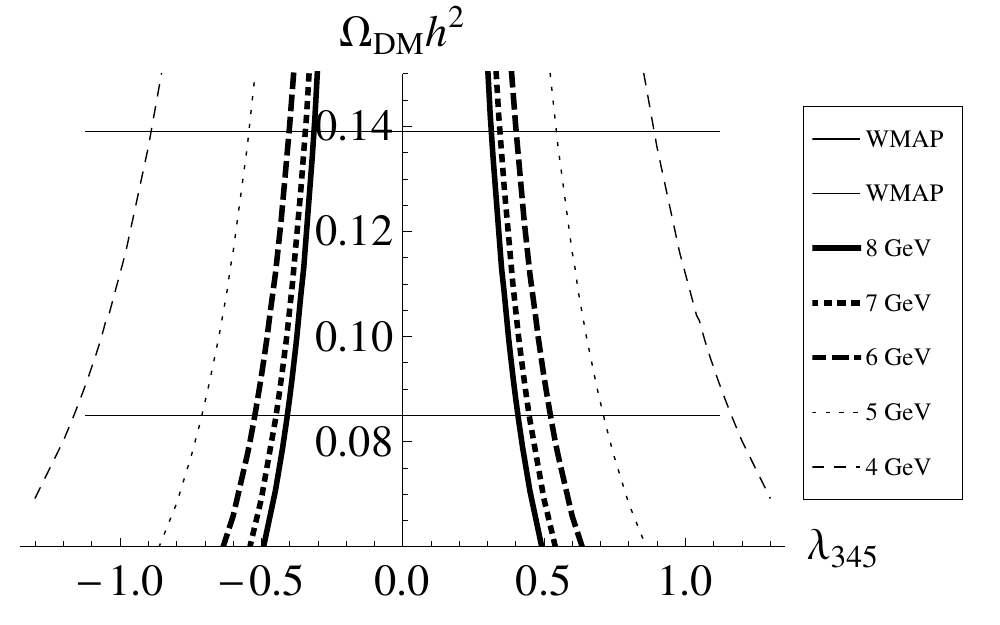} 
\vspace{-5pt}
  \caption{Relic density for $M_{D_H}=(4-8)$ GeV, $\lambda_2 = 5, \; \delta_A=100 \textrm{ GeV}, \; \delta_\pm=105$ GeV. Horizontal lines denote WMAP 3$\sigma$ allowed region. \label{omega4}}
\end{figure}

Large mass splittings between the $D_H$ and other scalar particles do not allow for the coannihilation. The main decay channel is $D_H D_H \to f \bar{f}$ ($c\bar{c}$ pair for $M_{D_H} = 4$ GeV and $b\bar{b}$ pair for higher $M_{D_H}$) via Higgs boson exchange. Small $|\lambda_{345}|$ generally gives high $\Omega_{DM} h^2$  well above the WMAP limit, as this decay channel is suppressed. The larger $|\lambda_{345}|$ is, the lower $\Omega_{DM} h^2$ gets. The WMAP allowed region of $|\lambda_{345}|$ is around $(0.4,1.2)$, but this strongly depends on the exact value of $M_{D_H}$ (figure \ref{omega4}). Note, that the 1 GeV difference in mass is this region causes significant change in the $\Omega_{DM} h^2$ for a given $\lambda_{345}$.

One should remember that changing of $\lambda_{345}$ is allowed only in the region allowed by the value of $\lambda_2$ (in the calculation of $\Omega_{DM} h^2$ we set $\lambda_2 = 5$ which corresponds to $1.5> \lambda_{345}>-1.5$). In general, as we allow the higher values of $\lambda_2$ the range of $\lambda_{345}$, over which we can scan, extends and for a chosen $M_{D_H}$ the higher $|\lambda_{345}|$ is the lower $\Omega_{DM} h^2$ we get.

For the low mass region relic density does not depend on the value of masses of the much heavier $D_A$ and $D^\pm$. In our analysis
we keep the mass splittings fixed to $\delta_A = 100, \delta_{\pm} = 105 \textrm{ GeV}$ and $M_{D_H}$ changes in the allowed region $(4-8)$ GeV. As $M_{D_H}$ decreases, the value of relic density for the chosen $\lambda_{345}$ grows, as shown in the figure \ref{omega4}.

\subsection{Medium DM mass region}

In the medium mass region the mass of $D_H$ is in range $M_{D_H} = (30-80) \textrm{ GeV}$. Mass splittings $\delta_A, \delta_{\pm}$ can be of the same order $\delta_A \approx \delta_{\pm} = (50-90) \textrm { GeV}$, but also small values of $\delta_A$ are possible (of the order of $10$ GeV) \cite{Dolle:2009fn}.

The medium mass region $\Omega_{DM} h^2$ is very sensitive to the value of $M_{D_H}$. Therefore in this case it is difficult to make a general statement similar to the low  mass case (section \ref{low_DM_sec}). However, we see some regularities in this behaviour, related mostly with the effects of coannihilation, as discussed below. For our calculation we use $\lambda_2 = 0.3$, which in this mass range allows scanning over $\lambda_{345}$ in range $\sim (-0.3,-0.35)$.

\paragraph{Small $\delta_A$}

\begin{figure}[htb]
\vspace{-10pt}
  \centering
  \subfloat[$\delta_A = 8$ GeV, $\delta_{\pm} = 50$ GeV]{\label{omega11}\includegraphics[width=0.4\textwidth]{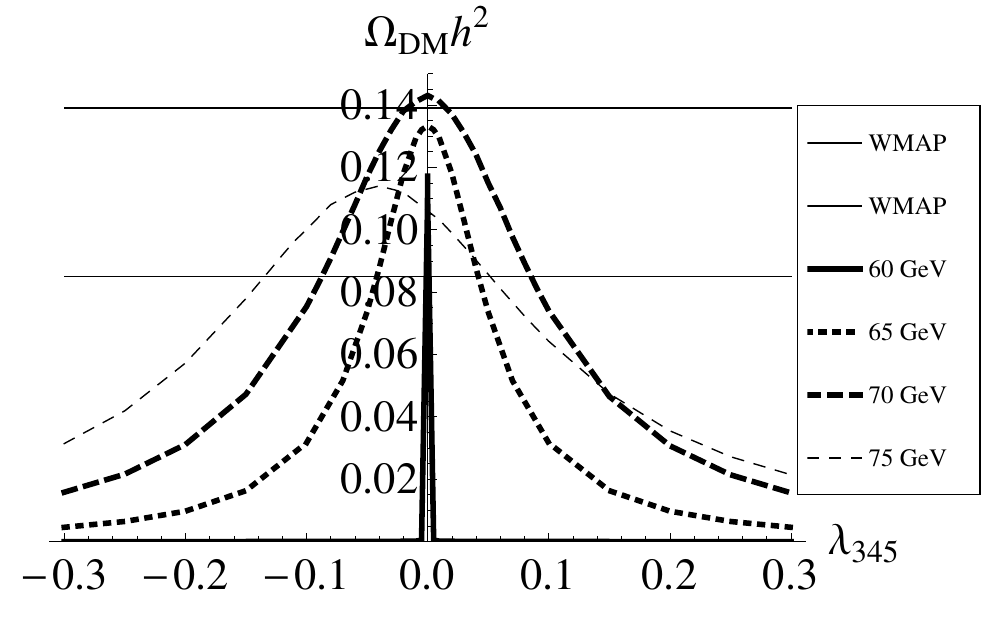}} \qquad
  \subfloat[$\delta_A = \delta_{\pm} = 70 \textrm{ GeV}$]{\label{omega16}\includegraphics[width=0.4\textwidth]{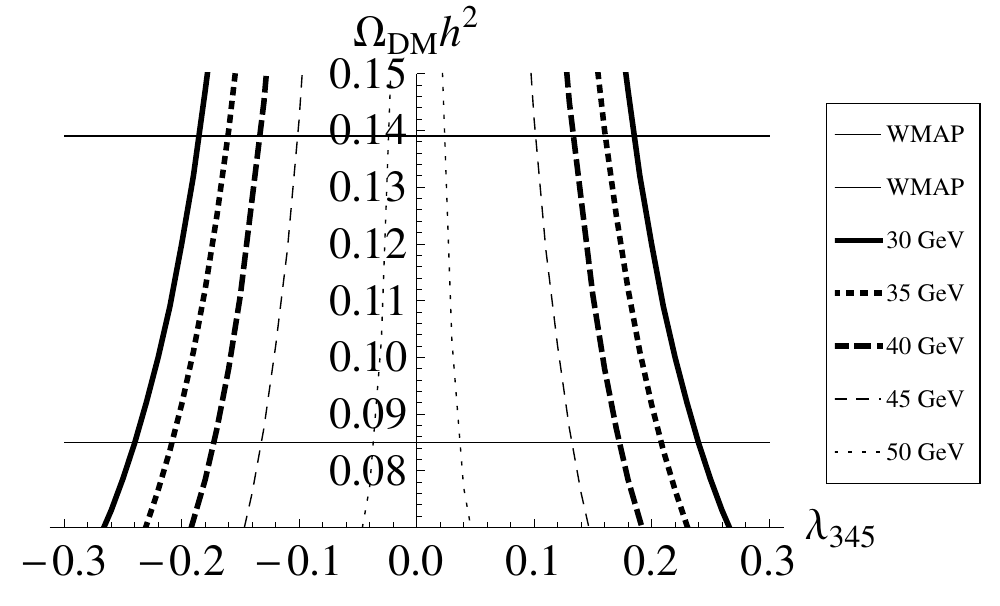}}
\vspace{-5pt}
  \caption{Relic density for medium mass region, results for $\lambda_2=0.3$. \label{omegamid_1}}
\end{figure}

Let us first consider the small mass splitting between $D_H$ and $D_A$ (figure \ref{omega11}):
\[\delta_A = 8 \textrm{ GeV }, \quad \delta_{H^\pm} = 50 \textrm{ GeV }.\]
Small $\delta_A$ makes the coannihilation $(D_H,D_A)$ important, decreasing the $\Omega_{DM} h^2$ below the WMAP limit in most of the $\lambda_{345}$ parameter space. Usually only the  region close to $\lambda_{345} = 0$ is inside the WAMP limit. 
For the larger masses of $D_H$, larger $\lambda_{345}$ are allowed. 

Note, that the smaller masses of $D_H$ which may give a different results are heavily constrained by the LEP II data.

\paragraph{Large $\delta_A$} 

If mass splittings between $D_H$ and other scalars are larger ($\delta \sim 50$ GeV) then the coannihilation is no longer important. 
 For smaller masses and large mass splittings the behaviour is similar to the one in the low mass region ($\lambda_{345}$ region around 0 is excluded and larger $|\lambda_{345}|$ are allowed, figure \ref{omega16}). The allowed values of $|\lambda_{345}|$ are larger than in the case of small $\delta_A$. 
 
 This region of masses is the most natural to consider as the constrains for the masses of scalars are not so tight as in other cases.

\subsection{High DM mass region}

In the high mass region $M_{D_H} = (500-1000) \textrm{ GeV}$ all dark scalars have almost degenerate masses and the mass splittings are small of the order of $\delta \approx \delta_A \approx \delta_{\pm} < 12 \textrm{ GeV}$ due to the perturbativity conditions \cite{Dolle:2009fn}. 

\begin{figure}[htb]
\centering
\includegraphics[scale=0.75]{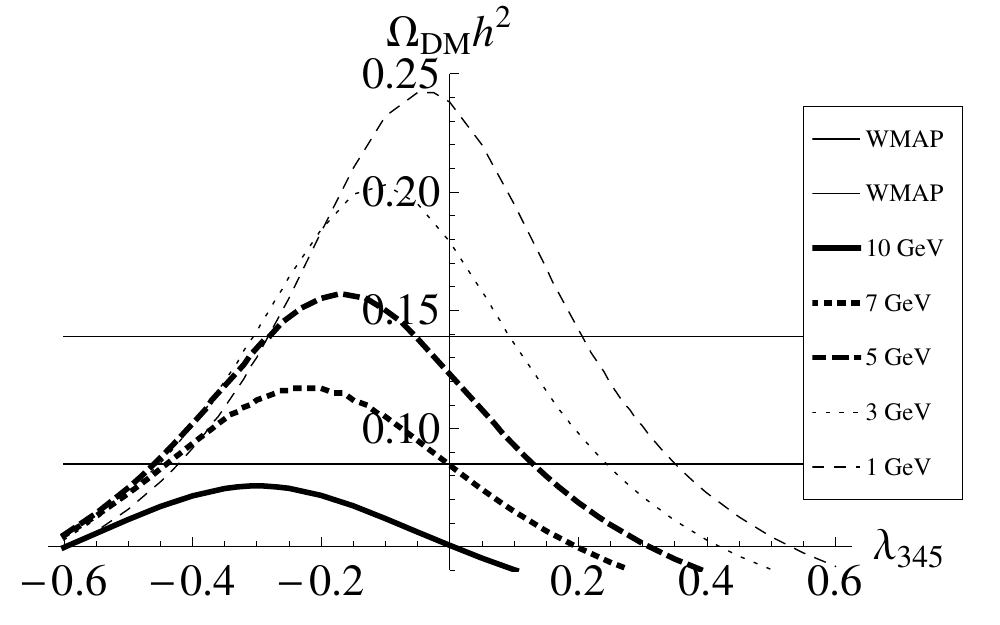}
  \caption{Relic density for high mass region: $(\lambda_{345},\Omega_{DM} h^2)$ plot for fixed DM mass $M_{D_H}=800$ GeV and $\delta_{A,\pm} =( 1,3,5,7,10)$ GeV, results for $\lambda_2=1.5$.}
  \label{omega8}
\end{figure}

In this high mass region, with such small mass splitting, the coannhilation between all dark scalars particles is very important. The results show very high sensitivity to the value of $\delta$'s, as shown in figure \ref{omega8}.
Here we fix $M_{D_H} = 800$ GeV, while $\delta$ varies from 1 to 10 GeV, respectively. In general, as $\delta$ grows, the value of relic density for given $\lambda_{345}$ decreases. For $\delta = 10$ GeV, for every value of $\lambda_{345}$, we are below the WMAP limit. 

\begin{figure}[htb]
\vspace{-10pt}
  \centering
  \subfloat[$(\lambda_{345},\Omega_{DM} h^2)$ for $\delta = 1$ GeV]{\label{omega9}\includegraphics[width=0.4\textwidth]{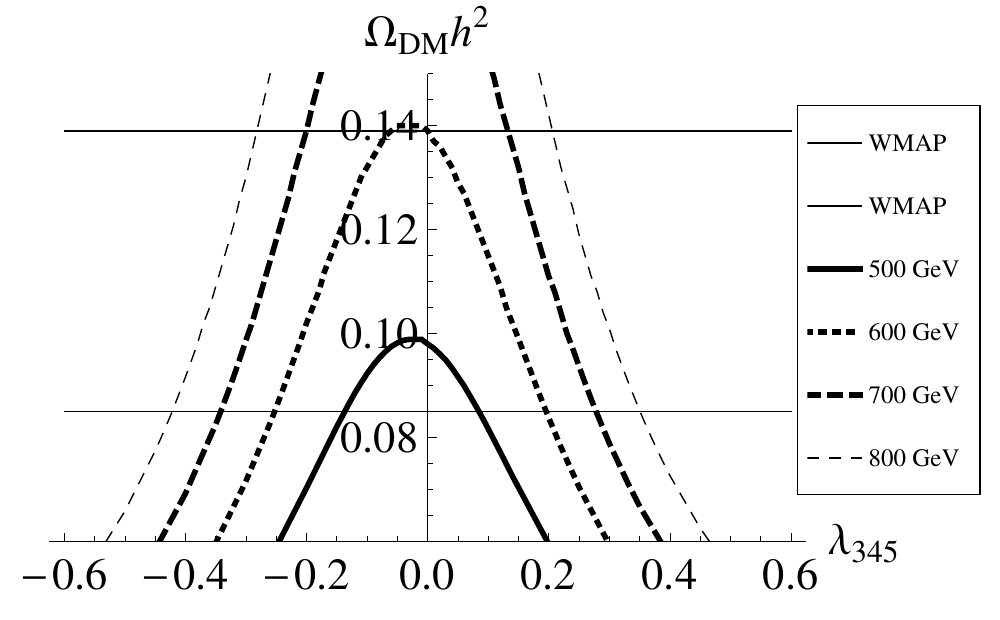}} \qquad
  \subfloat[$(\lambda_{345},\Omega_{DM} h^2)$ for $\delta = 10$ GeV]{\label{omega10}\includegraphics[width=0.4\textwidth]{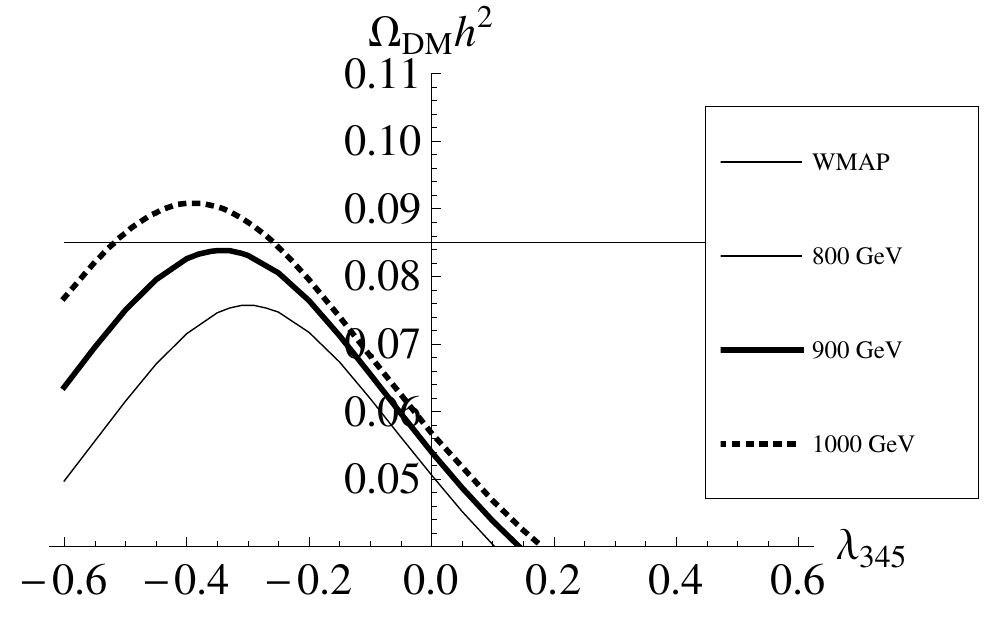}}
\vspace{-5pt}
  \caption{Relic density for high mass region for varied DM mass. Results for $\lambda_2 =1.5$. \label{omegahigh}}
\end{figure}

In figure \ref{omegahigh} $M_{D_H}$ changes between $500$ and $1000 \textrm{ GeV}$, while $\delta$'s are fixed. We first consider $\delta = 1$ GeV (figure \ref{omega9}), where lower values of $M_{D_H}$ correspond to the lower values of $\Omega_{DM} h^2$. 
If $\delta = 10$ GeV (figure \ref{omega10}) then for most of the chosen values of $M_{D_H}$ we are below the WMAP limit. Mass around 1000 GeV gives the proper relic density for higher values of $|\lambda_{345}|$, note that $\lambda_{345}<0$.

\section{Thermal evolution}

\subsection{Thermal evolution of the Universe}
\paragraph{Evolution of the potential} Following the approach presented in \cite{Ginzburg:2007jn, Ginzburg:2009dp, Ivanov:2008er, Ginzburg:2010wa,nasza_druga} we consider the first nontrivial temperature corrections to the potential $V$. 
The coefficients $\lambda's$ of  the quartic terms  are unchanged, while the quadratic terms vary with temperature $T$ with coefficients:
\bear{c}
m_{11}^2(T)=  m_{11}^2-c_1T^2\,,\quad
m_{22}^2(T) =  m_{22}^2- c_2T^2\,,\\[3mm]
c_1=\fr{3\lambda_1+2\lambda_3+\lambda_4}{6}+\fr{3g^2+g^{\prime 2}}{8}+\fr{g_t^2+g_b^2}{2},\quad 
c_2=\fr{3\lambda_2+2\lambda_3+\lambda_4}{6}+\fr{3g^2+g^{\prime 2}}{8}.			 \eear{Tempdep}

The EW gauge couplings $g$ and $g^\prime$ and Yukawa couplings for $b$ and $t$ quarks are:

\begin{equation}
g= 2 M_W /v = 0.652, \; g'=0.351; \; g_i = \sqrt{2} m_i/v \; (g_t \approx 0.99, g_b \approx 0.02).
\end{equation}

As shown in \cite{Ginzburg:2010wa}, depending on the value of $\lambda's$ each of the coefficients $c_1$ and $c_2$ can be either positive or negative. In this work we don't discuss the possibility of non-restoration of the EW symmetry in the past \cite{Gavela:1998ux} and therefore we consider  $c_2,c_1>0$ only \cite{Ginzburg:2010wa}.

\paragraph{Extrema during the evolution} Depending on the values of the parameters the $D$-symmetric potential (\ref{pot}) may have different types of the vacuum state. The general form of an neutral extremum is given by the following solution of the extremum conditions: 
\bear{c}
        \langle\Phi_S\rangle =\dfrac{1}{\sqrt{2}}\left(\begin{array}{c} 0\\
        v_S\end{array}\right),\quad \langle\Phi_D\rangle
        =\dfrac{1}{\sqrt{2}}\left(\begin{array}{c} 0 \\ v_D
        \end{array}\right),\quad (v^2=v_S^2+v_D^2).
\eear{genvac}

The properties of these extrema, provided they are realized as vacua, are listed in the table \ref{vacua}. As we discussed in \cite{Ginzburg:2010wa}, each of this vacua can be realized in the separate regions in $(\mu_1,\mu_2)$ plane. 

 \begin{table}
{\renewcommand{\arraystretch}{1.5}

\begin{tabular}{|c|c|p{7.5cm}|}
\hline
\hline
\multicolumn{3}{|c|}{Extrema and vacua} \\ \hline \hline
name of extremum & vev's &properties of vacuum\\\hline
\textit{EW symmetric}: $EW\! s$ & $v_D=0, \quad v_S=0$ & Massless fermions and bosons and massive scalar doublets. \\ \hline
 \textit{inert}: $I_1$ & $v_D=0,\quad v_S^2=v^2=\fr{m_{11}^2}{\lambda_1}$ & Massive fermions and gauge bosons; scalar sector: SM-like Higgs $h_S$ and dark scalars $D_H, D_A, D^\pm$ with \textbf{DM candidate} $D_H$. \\ \hline
\textit{inertlike}: $I_2$ & $v_S=0,\quad v_D^2=v^2=\fr{m_{22}^2}{\lambda_2}$ & Massless fermions and massive gauge bosons; scalar sector: Higgs particle $h_D$ (no interaction with fermions), four scalars $S_H,S_A,S^\pm$, \textbf{no DM candidate}. \\ \hline
\textit{mixed}: $M$ & $v_S^2, v_D^2 > 0, \quad v^2=v_S^2+v_D^2$ & Massive fermions and bosons, 5 Higgs particles: CP-even $h$ and $H$, CP-odd $A$ and charged $H^\pm$, \textbf{no DM candidate}. \\ \hline
\hline
\end{tabular}
}
  \caption{Extrema and properties of the vacua \cite{Ginzburg:2010wa}. \label{vacua}}
\end{table}

\begin{figure}[hb]
\vspace{-10pt}
  \centering
  \subfloat[$R>1$]{\label{fig:wykresa}\includegraphics[width=0.3\textwidth]{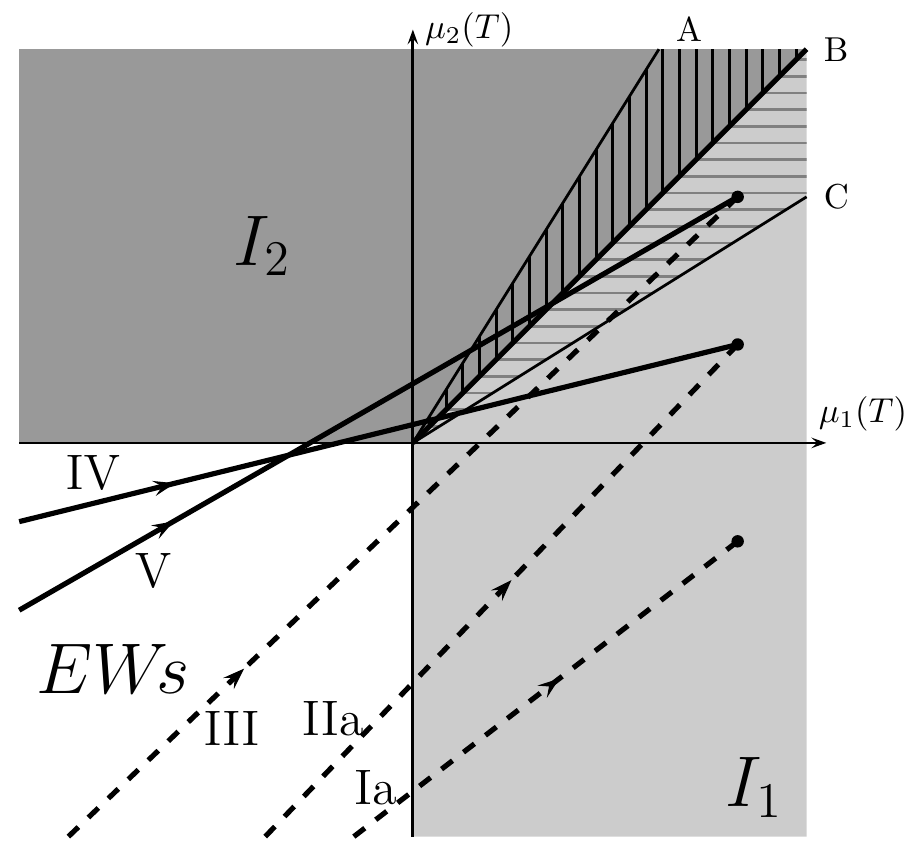}} \qquad
  \subfloat[$1>R>0$]{\label{fig:wykresb}\includegraphics[width=0.3\textwidth]{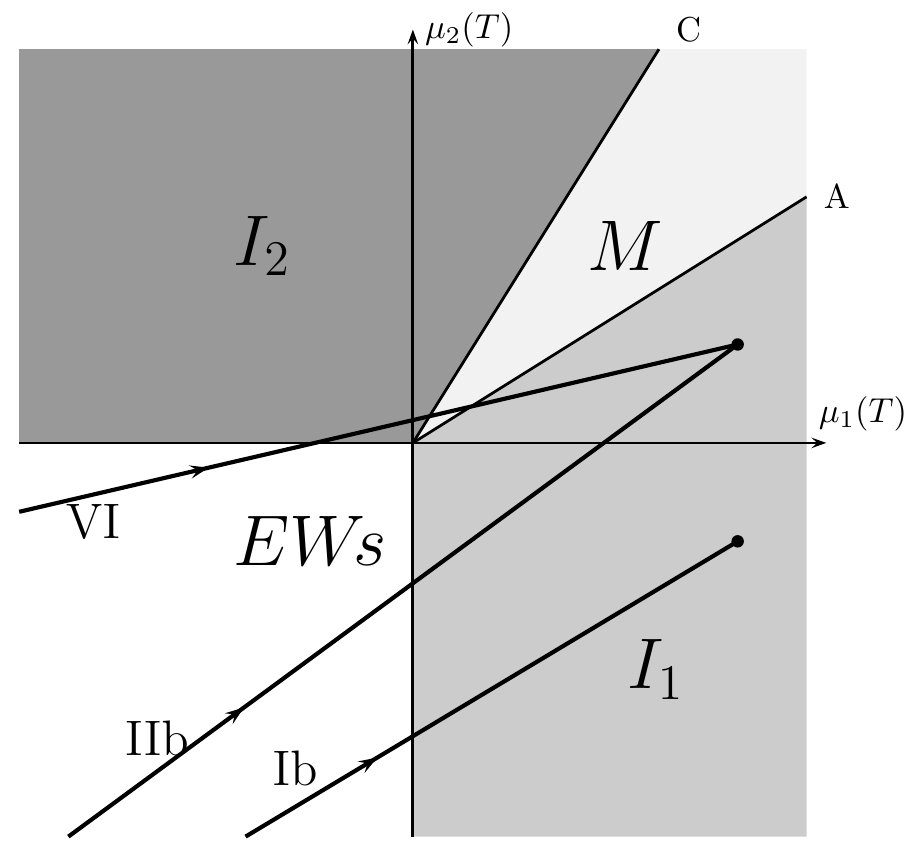}} \qquad
  \subfloat[$0>R>-1$]{\label{fig:wykresc}\includegraphics[width=0.3\textwidth]{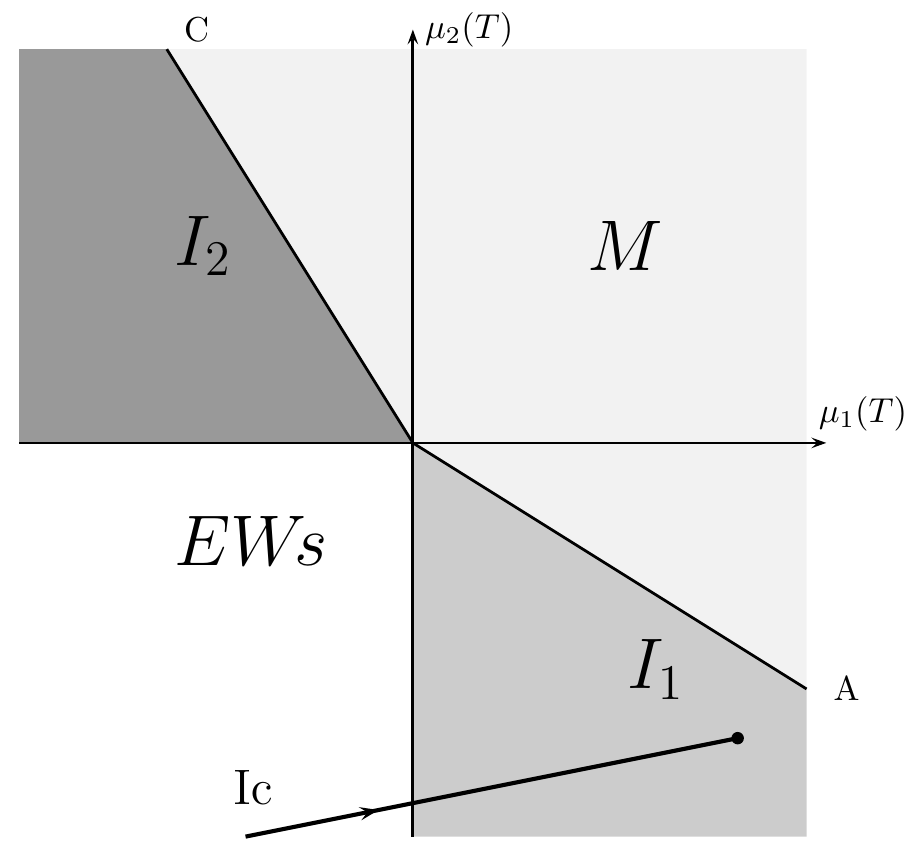}}
\vspace{-5pt}
  \caption{Possible rays on ($\mu_1,\mu_2$) plane. Vacua: EW symmetric (white region), inert (yellow, lightest shade), inertlike (blue, dark shade) and mixed (green, medium shade). Vertical hatch between lines $A$ and $B$ in figure \ref{fig:wykresa} denotes the area when $I_1$ is a local minimum, while $I_2$ is the global one; horizontal hatch between $B$ and $C$ -- $I_2$ is a local minimum, while $I_1$ is the vacuum. }
  \label{fig:vacua}
\end{figure}

\paragraph{Types of evolution} As the quadratic terms in $V$ vary with $T$, the vacuum state changes. As time grows (and temperature decreases) the vacuum state of the Universe changes along one of rays in the phase diagrams shown in figure \ref{fig:vacua}, where the allowed regions for all vacua $EW\! s$, $I_1$, $I_2$ and $M$ are shown as well.

The types of evolution (rays), along with conditions on $R$ and $\mu_1, \mu_2$ that need to be satisfied in order to have a chosen type of evolution, are summarised in the table \ref{raym1m2} in the appendix. Note, that in existing in literature DM analysis only the rays that correspond to a single phase transition $EW\! s \to I_1$ have been considered.

\subsection{Rays in the $(\lambda_{345},\lambda_2)$ plane}

To get a deeper insight into evolution of the Universe we consider now parametrization of rays by the physical parameters like masses and self-couplings. We will use the $(\lambda_{345},\lambda_2)$ plane, to represent regions where a certain ray is realized, treating masses of scalars $M_{h_s}, \, M_{D_H}, \, M_{D_A}, \, M_{D^\pm}$ as the input parameters. Conditions from table \ref{raym1m2} have been translated into conditions for $(\lambda_{345},\lambda_2)$ parameters and are presented in table \ref{raylambda} in the appendix. Note, that only for the case of a single phase transition we need to impose the constraints for $\lambda_{345}$ parameter. In case of sequences with two or three phase transitions, constraints on the quartic coupling $\lambda_2$ are sufficient.

The $(\lambda_{345},\lambda_2)$  phase space can be divided into separate sectors. 
Let us introduce $\tilde \lambda_{345}$ parameter, which separates these sectors:
\begin{equation}
\tilde \lambda_{345} = \frac{2 M_{D_H}^2}{v^2}.
\end{equation}

For $\lambda_{345}<\tilde \lambda_{345}$, which corresponds to $\mu_2 < 0$, the only $EW\! v$ extremum, and thus the vacuum, that ever existed is the inert phase $I_1$.  The only possible sequence is $EW\! s \to I_1$.
For $\lambda_{345}>\tilde \lambda_{345}$ ($\mu_2 > 0$) other $EW\! v$ extrema ($M,I_2$) could be realized in the past. Their realization as vacua depends on the value of $\lambda_2$.

Distinct regions of $\lambda_2$ are separated by  $\tilde \lambda_2$, which depends on $\lambda_{345}$ and other parameters as follows:
\begin{eqnarray}
\tilde \lambda_2 &=&  \frac{-(3 g^2 + g^{\prime 2})}{4}- \frac{ (M_{D_A}^2 - 3 M_{D_H}^2 + 2 M^2_{D^\pm})}{3 v^2} + \frac{\lambda_{345}}{3}  + \frac{(2 M_{D_H}^2 - v^2 \lambda_{345})}{12} \times  \nonumber \\ &&  \times \frac{(-4 M_{D_A}^2 + 12 M_{D_H}^2 - 8 M^2_{D^\pm} - (9 g^2  + 3 g^{\prime 2}  + 12 g_b^2  + 12 g_t^2  + 8  \lambda_{345})v^2)}{v^2 M_{h_s}^2 }.
 \label{l2bond}
\end{eqnarray}
Note, that for $\lambda_2 > \tilde \lambda_2$ only one phase transition may be realized, while for $\lambda_2 < \tilde \lambda_2$ sequences with two or three phase transitions are possible.

Below we perform a general analysis of possible rays on the $(\lambda_{345},\lambda_2)$ plane for various masses, assuming that today $I_1$ vacuum is realized. 

\subsection{Benchmark points} 

For further discussions we will use the
reference points $B1-B4$ chosen in three regions of $M_{D_H}$ as follows. For all points $M_{h_S} = 120$ GeV. 
\begin{itemize}
\item Low DM mass region: for the reference point $B1$ we have:
\begin{equation}
B1 \textrm{ : } \quad M_{D_H} = 8 \textrm{ GeV}, \; \delta_A = 100 \textrm{ GeV}, \; \delta_{\pm} = 105 \textrm{ GeV}. \label{B1}
\end{equation}
\item Medium DM mass region: as discussed, large or small $\delta_A$ are possible. Those are the benchmark points $B2$ and $B3$, respectively.
\begin{equation}
B2 \textrm{ : } \quad M_{D_H} = 45 \textrm{ GeV}, \; \delta_A = 70 \textrm{ GeV}, \; \delta_{H^\pm} = 70 \textrm{ GeV}, \label{B2}
\end{equation}
\begin{equation}
B3 \textrm{ : } \quad M_{D_H} = 70 \textrm{ GeV},  \; \delta_A = 8 \textrm{ GeV}, \; \delta_{H^\pm} = 50 \textrm{ GeV}. \label{B3}
\end{equation}
\item  The high DM mass region  set $B4$ is defined as:
\begin{equation}
B4 \textrm{ : } \quad M_{D_H} = 800 \textrm{ GeV}, \; \delta_A = 1 \textrm{ GeV}, \; \delta_{H^\pm} = 1 \textrm{ GeV}. \label{B4}
\end{equation}
\end{itemize}

\begin{figure}[htb]
\vspace{-10pt}
  \centering
  \subfloat[$B1 = (8;100,105)$]{\label{low1}\includegraphics[width=0.22\textwidth]{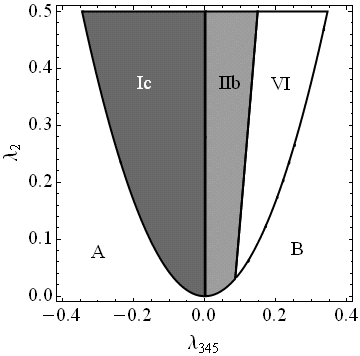}} \;
  \subfloat[$B2=(45;70,80)$]{\label{mid11}\includegraphics[width=0.22\textwidth]{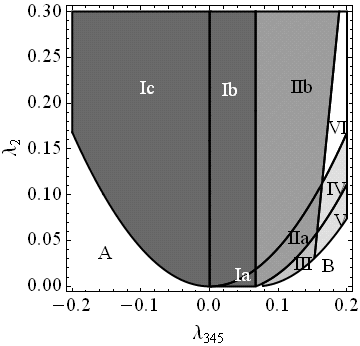}} \;
  \subfloat[$B3=(70;8,50)$]{\label{mid21}\includegraphics[width=0.22\textwidth]{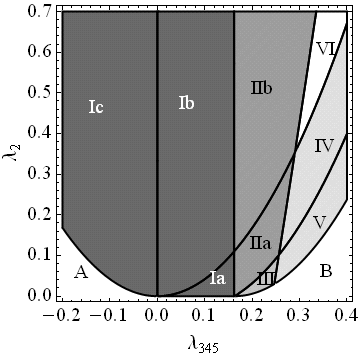}} \;
  \subfloat[$B4=(800;1,1)$]{\label{high1}\includegraphics[width=0.22\textwidth]{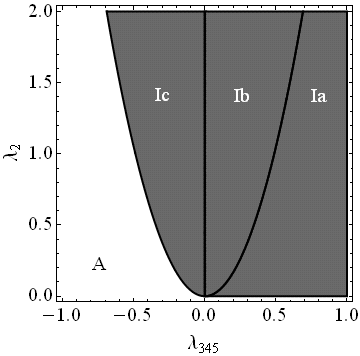}}
\vspace{-5pt}
  \caption{Regions of rays for reference points $Bi=(M_{D_H};\delta_A,\delta_\pm) \textrm{ GeV}$ for $M_{h_S} = 120 \textrm{ GeV}$. Different colours correspond to the different rays. $EW\! s \to I_1$: Ia, Ib, Ib, IIa, IIb, III; $EW\! s \to I_2 \to I_1$: IV and V; $EW\! s \to I_2 \to M \to I_1$: VI. Region A excluded by positivity constraints, region B excluded by $I_2$ vacuum.  \label{fig:baseplots}}
\end{figure}

Figures (\ref{fig:baseplots}a-d) show the regions, where concrete rays can be realized in $(\lambda_{345},\lambda_2)$ plane for mass points $B1-B4$.\footnote{Note, that the scales in those figures are different.} In general, the higher $D_H$ mass is, the higher values of $\lambda_{345}$ are needed for the realization of the sequences with more than one phase transition. More discussion is given below (sec. \ref{ben_low}-\ref{ben_high}), where also the relic density as the function of $\lambda_{345}$ is discussed.

\subsection{Low DM mass region \label{ben_low}}

\paragraph{Rays in $(\lambda_{345},\lambda_2)$ plane} Figure \ref{low1} shows the $(\lambda_{345},\lambda_2)$ plane for the low mass point $B1$. In this case possible types of evolution are very limited. There is an allowed region for the single phase transition (realized by ray Ic for $R<0$ and ray IIb for $1>R>0$) and sequence of three phase transitions with transition through $M$ (ray VI). Important feature of this low DM mass region is the fact that 1st-order phase transition $EW\! s \to I_2 \to I_1$ is not possible (rays IV and V), as the allowed phase space for those rays shrinks to the degenerate line $\lambda_2 \approx \lambda_{345}^2 v^2/M_{h_s}^2$. In short, fulfilling $R>1$ condition is not possible for the low DM mass region.

\begin{figure}[hb]
\vspace{-10pt}
  \centering
  \subfloat[$(\lambda_{345},\Omega_{DM} h^2)$ for B1]{\label{omega1}\includegraphics[width=0.3\textwidth]{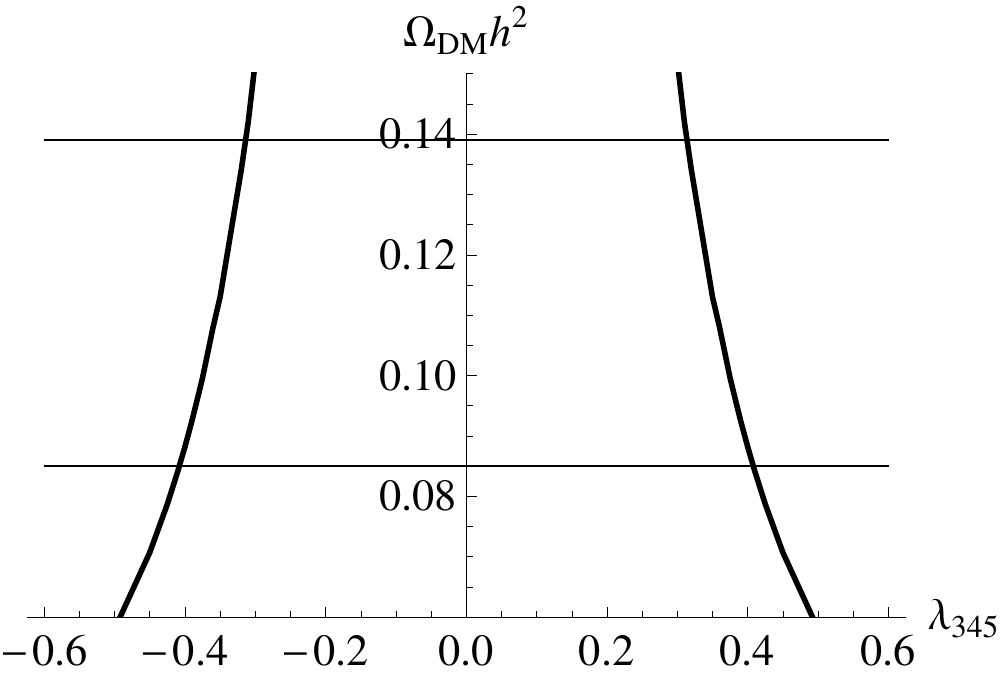}} \qquad
  \subfloat[$(\lambda_{345},\lambda_{2})$ for B1]{\label{omega2}\includegraphics[width=0.2\textwidth]{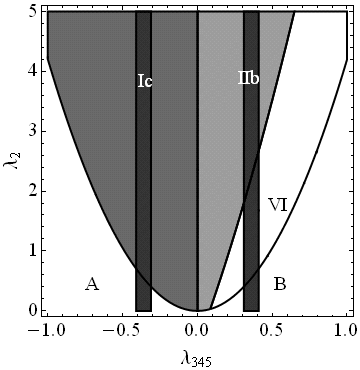}}
\vspace{-5pt}
  \caption{Relic density for $B1$ with $M_{D_H} = 8 \textrm{ GeV}, \; \delta_A = 100 \textrm{ GeV}, \; \delta_\pm = 105 \textrm{ GeV}$. \label{omegalow}}
\end{figure}

\paragraph{Relic density} The obtained energy relic density as a function of $\lambda_{345}$ is presented in 
figure \ref{omega1} for a low mass benchmark point B1. There are two regions in agreement with the WMAP limit: $\lambda_{345} \in ( -0.405, -0.315)$ and $\lambda_{345} \in ( 0.315, 0.405)$, presented as the dark regions in $(\lambda_2,\lambda_{345})$ plot (figure \ref{omega2}). For the negative values of $\lambda_{345}$ ray Ic, for positive -- ray VI or ray IIb, depending on the value of $\lambda_2$, can be realized. Note, that the value of $\lambda_2$ should be quite large to have any kind of solution that fits into the WMAP region. Because of the positivity constraints for $\lambda_{345} \in ( -0.405, -0.315)$ we need $\lambda_2 > 0.45$ if we want to have ray Ic.\footnote{For positive allowed $\lambda_{345}$ and $\lambda_2 < 0.45$ the Universe is in the state of intertlike vacuum.} For $\lambda_{345}>0$ and $1.9 > \lambda_2 > 0.45$ only ray VI is realized; if we want to have ray IIb we need $\lambda_2 > 1.9$.

Calculation of $\Omega_{DM} h^2$ was done for $\lambda_2 = 5$, which does not influence its value. However, as we discussed above, the value of $\lambda_2$ is  limited by the other constrains for certain rays. 
It is limited by the positivity constraints $\lambda_{345} + \sqrt{\lambda_1 \lambda_2} >0$ for $\lambda_{345}<0$, while for $\lambda_{345}>0$ positivity constraints don't give the further constraints on $\lambda_2$, but its value is limited by the conditions of realization of particular rays (table \ref{raylambda}).

\subsection{Medium mass region \label{ben_mid}}

\paragraph{Rays in $(\lambda_{345},\lambda_2)$ plane} Figure \ref{mid11} shows the allowed rays for the reference point $B2$ in the medium mass region for $M_{D_H} = 45$ GeV. In this case all types of evolution are possible. Note, that rays that correspond to the sequence of two or more phase transitions (rays IV, V and VI) appear for larger values of $\lambda_{345}$. For the point $B3$, with $M_{D_H} = 70$ GeV, the types of rays are similar to the $B2$ case, as shown in figure \ref{mid21}. Some small differences are visible, for example ray VI requires larger value of $\lambda_2$. Those differences are mainly due to the value of $M_{D_H}$, mass splittings $\delta_{A},\delta_\pm$ are less important (see section 5).

\begin{figure}[htb]
\vspace{-10pt}
  \centering
  \subfloat[$(\lambda_{345},\Omega_{DM} h^2)$ for B2 ]{\label{omega12a}\includegraphics[width=0.35\textwidth]{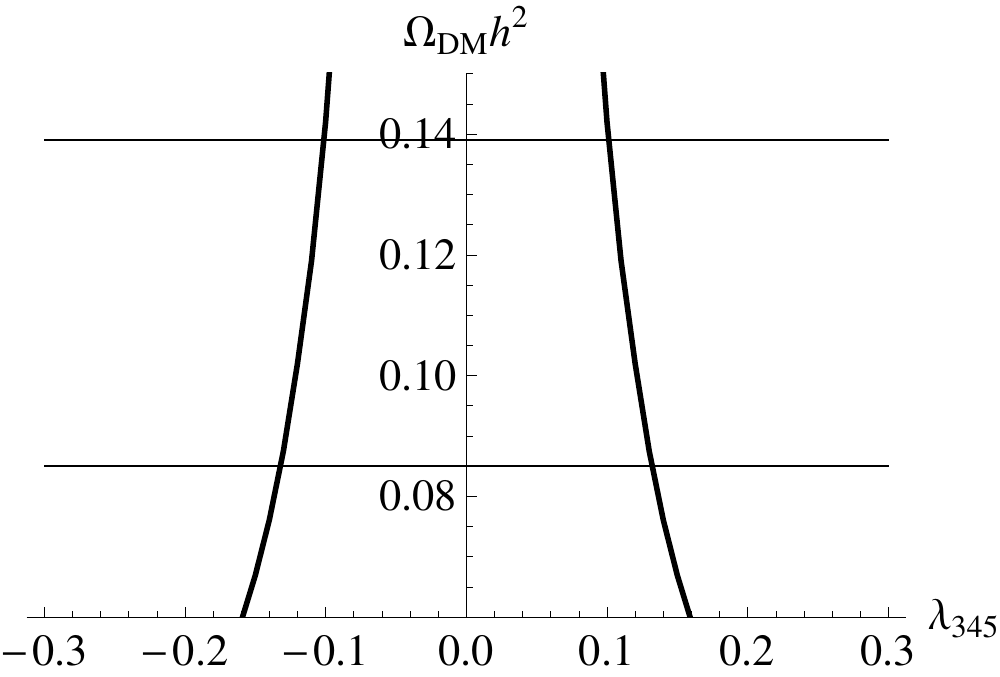}} \qquad
  \subfloat[$(\lambda_{345},\lambda_2)$ for B2]{\label{omega13a}\includegraphics[width=0.25\textwidth]{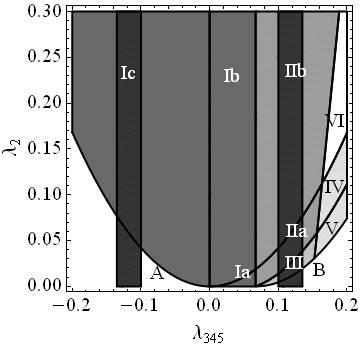}}
\vspace{-5pt}
  \caption{Relic density  for $B2$ with $M_{D_H} = 45 \textrm{ GeV},\;\delta_A=\delta_\pm = 70 \textrm{ GeV}$.}
  \label{fig:middle2}
\end{figure}

\paragraph{Relic density} Point $B2$ is similar to the low mass region as here the coannihilation is not relevant. However, due to the larger $D_H$ mass the annihilation through Higgs exchange is more efficient and the obtained $\lambda_{345}$ values are smaller. Again there are two allowed regions that give the proper relic density (figure \ref{fig:middle2}). Note, that this time for the positive $\lambda_{345}$ rays IIb, IIa and III (the later with the coexisting local minimum $I_2$) are  allowed by the relic density data.

\begin{figure}[htb]
\vspace{-10pt}
  \centering
  \subfloat[$(\lambda_{345},\Omega_{DM} h^2)$ for B3]{\label{omega12}\includegraphics[width=0.35\textwidth]{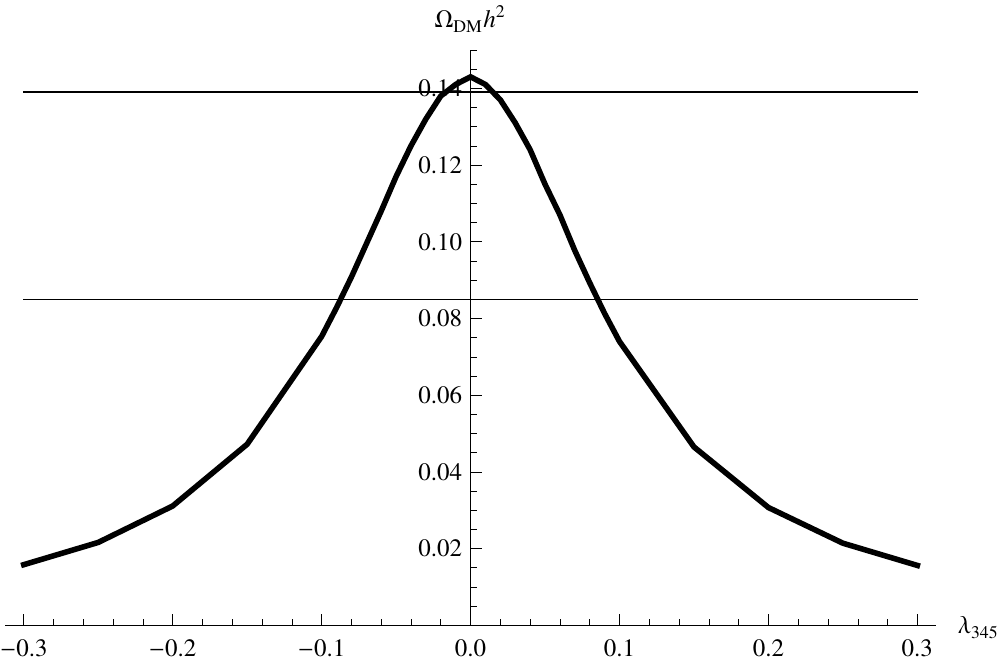}} \qquad
  \subfloat[$(\lambda_{345},\lambda_2)$ for B3 ]{\label{omega13}\includegraphics[width=0.25\textwidth]{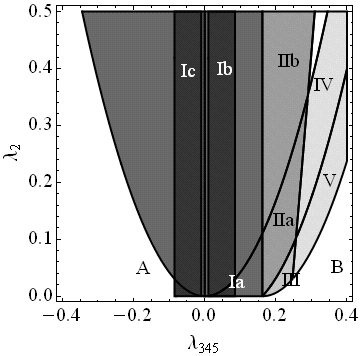}}
\vspace{-5pt}
  \caption{Relic density  for $B3$ with $M_{D_H} = 70 \textrm{ GeV}, \delta_A = 8 \textrm{ GeV}, \; \delta_\pm = 50 \textrm{ GeV}$.}
  \label{fig:middle}
\end{figure}

 In case of the point $B3$ due to the small value of $\delta_A$ the coannihilation $(D_H,D_A)$ is important for the relic density. Its effects are clearly visible for the very small values of $\lambda_{345}$, where annihilation through Higgs is strongly suppressed (figure \ref{omega12}). We get the proper relic density for $\lambda_{345} \in (-0.1,0.1)$ with the small region around 0 excluded. This corresponds to rays Ia-c, depending on the value of $\lambda_2$ (figure \ref{omega13}). Here, sequences with more than one phase transition give $\Omega_{DM} h^2$ below the WMAP limit.

\subsection{High mass region \label{ben_high}}

\paragraph{Rays in $(\lambda_{345},\lambda_2)$ plane} For small values of $\lambda_{345}$ only rays Ic, Ib and Ia are possible (figure \ref{high1}). The types of evolution other than a single phase transition could happen only for the very large values of $\lambda_{345}>2M_{D_H}^2/v^2$, which for $M_{D_H} = O(800) \textrm{ GeV}$ gives $\lambda_{345} = O(20)$. 

\begin{figure}[hb]
\vspace{-10pt}
  \centering
  \subfloat[$(\lambda_{345},\Omega_{DM} h^2)$ for B4]{\label{omega5}\includegraphics[width=0.4\textwidth]{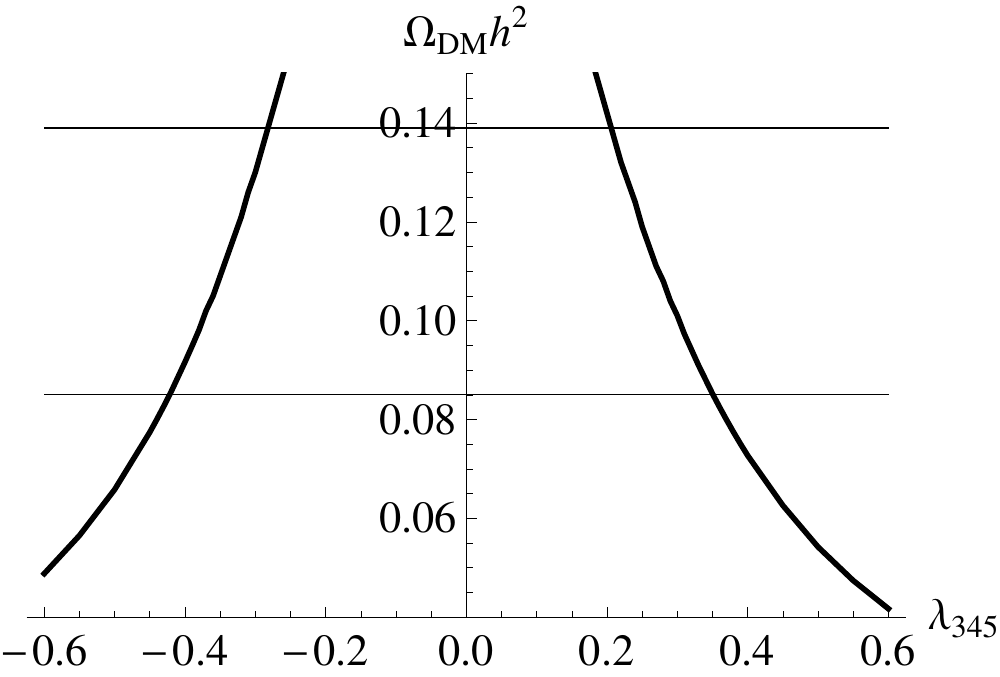}} \qquad
  \subfloat[$(\lambda_{345},\lambda_{2})$ for B4]{\label{omega6}\includegraphics[width=0.3\textwidth]{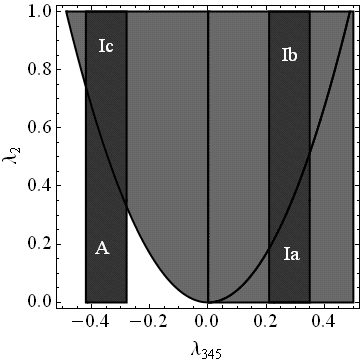}}
\vspace{-5pt}
  \caption{Relic density for $B4$ with $M_{D_H} = 800 \textrm{ GeV}, \delta_{A,\pm} = 1 \textrm{ GeV}$. \label{omegahigset}}
\end{figure}

\paragraph{Relic density} In high DM mass region, due to the very small mass splittings coannihilation processes are the most important ones. 
For chosen set of masses the allowed ranges of $\lambda_{345}$ are $(-0.42,-0.28),(0.21,0.35)$, as shown in figure \ref{omega5}.  It is clear that for this set the allowed rays are Ia, Ib, Ic (single phase transitions). The other rays could be realized for $\lambda_{3} > 22$, but first of all, it would violate the perturbativity constraints. On the other hand, the value of $\Omega_{DM} h^2$ is very low (of the order of $10^{-3}$) for such large $\lambda_3$ (and $\lambda_{245}$).

\section{Sensivity of the types of evolution to the dark scalar masses}

The realization of different types of evolution depends strongly on values of all parameters in the potential, as presented in numerical examples in \cite{nasza_druga}. However, in general the value of $M_{D_H}$ affects the type of evolution stronger than the values of $M_{D_A}$ and $M_{D^{\pm}}$. 

Let us consider first the values of masses for dark scalars allowed by the relic density constraints for the low and medium mass of $M_{D_H} \approx (8-80) \textrm{ GeV}$:
\begin{eqnarray}
M_{D_A} = 110 \textrm{ GeV}, \quad M_{D^\pm} = 110 \textrm{ GeV}.
\end{eqnarray}

\begin{figure}[hb]
\vspace{-10pt}
  \centering
  \subfloat[$M_{D_H} = 8 \textrm{ GeV}$]{\label{fig:s1}\includegraphics[width=0.23\textwidth]{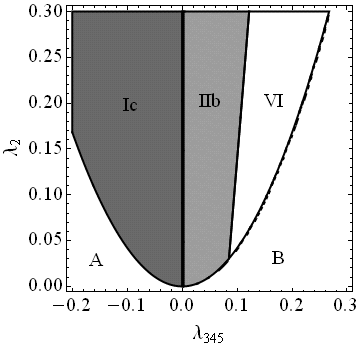}} \;
  \subfloat[$M_{D_H} = 35 \textrm{ GeV}$]{\label{fig:s3}\includegraphics[width=0.23\textwidth]{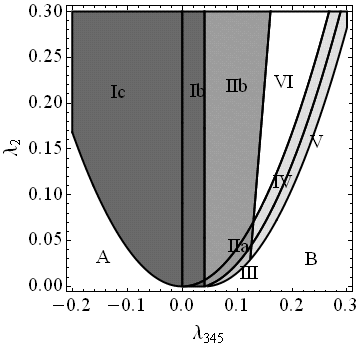}} \;
  \subfloat[$M_{D_H} = 45 \textrm{ GeV}$]{\label{fig:s4}\includegraphics[width=0.23\textwidth]{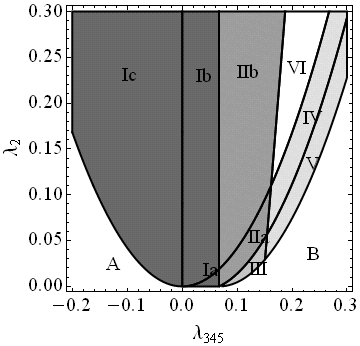}} \;
  \subfloat[$M_{D_H} = 60 \textrm{ GeV}$]{\label{fig:s6}\includegraphics[width=0.23\textwidth]{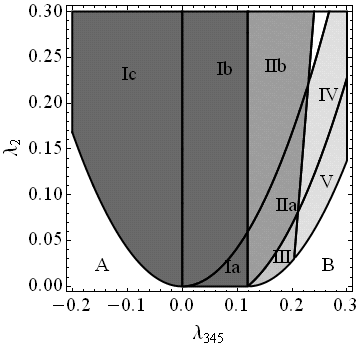}}
\vspace{-5pt}
  \caption{Regions of rays for $M_{h_S} = 120 \textrm{ GeV}, M_{D_A} = 110 \textrm{ GeV}, M_{D^\pm} = 110 \textrm{ GeV}$. $M_{D_H}$ varies. \label{fig:set1}}
\end{figure}

Figures (\ref{fig:set1}a-d) show the regions of parameters on $(\lambda_{345},\lambda_2)$ for the different values of $M_{D_H}$ (from $8$ GeV to $60$ GeV). For the low mass region single phase transition (rays Ic, IIb) and transition through $M$ vacuum (ray VI) are possible, other types of evolution appear for the higher values of $M_{D_H}$. As the mass grows, the region of ray IIa expands and thus ray VI becomes possible for the larger values of $\lambda_2$.

\begin{figure}[hb]
\vspace{-10pt}
  \centering
  \subfloat[$(50,60)$]{\label{fig:z1}\includegraphics[width=0.23\textwidth]{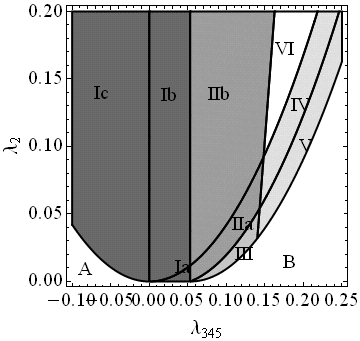}} \;
  \subfloat[$(60,75)$]{\label{fig:z2}\includegraphics[width=0.23\textwidth]{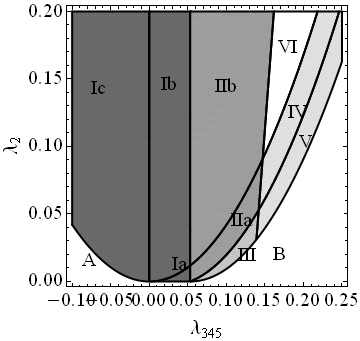}} \;
  \subfloat[$(70,80)$]{\label{fig:z3}\includegraphics[width=0.23\textwidth]{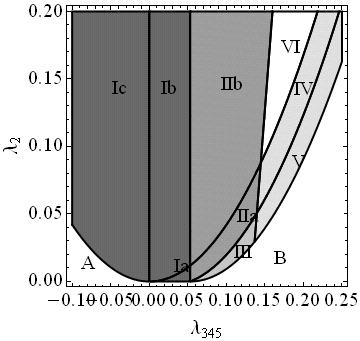}} \;
  \subfloat[$(80,80)$]{\label{fig:z4}\includegraphics[width=0.23\textwidth]{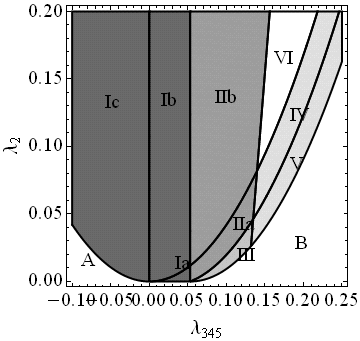}}
\vspace{-5pt}
  \caption{$M_{D_H} = 40 \textrm{ GeV}, M_{h_S} = 120 \textrm{ GeV}$. $M_{D_A}$ and $M_{D^\pm}$ vary. Cases (a-d) done for the pairs $(\delta_A, \delta_\pm)$ GeV. \label{fig:set2}}
\end{figure}

\begin{figure}[h]
\vspace{-10pt}
  \centering
  \subfloat[$M_{D_H} = 42 \textrm{ GeV}$]{\label{fig:mz1}\includegraphics[width=0.23\textwidth]{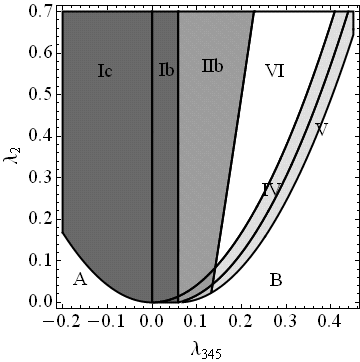}} \;
  \subfloat[$M_{D_H} = 60 \textrm{ GeV}$]{\label{fig:mz2}\includegraphics[width=0.23\textwidth]{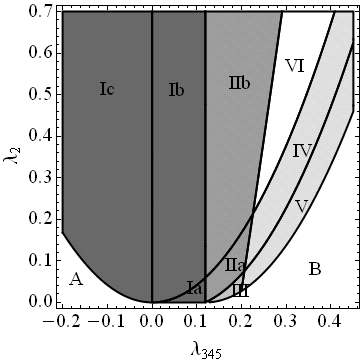}} \;
  \subfloat[$M_{D_H} = 80 \textrm{ GeV}$]{\label{fig:mz3}\includegraphics[width=0.23\textwidth]{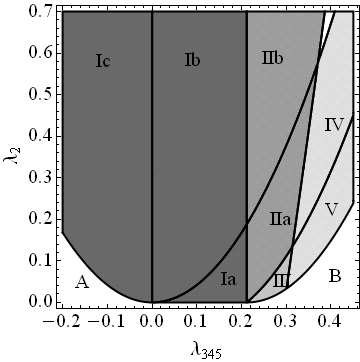}}
\vspace{-5pt}
\caption{Medium mass region for $\delta_A = 8 \textrm{ GeV}, \delta_\pm = 50 \textrm{ GeV} $. \label{fig:mid12} }
\end{figure}

If we fix the value of $M_{D_H}$ and let the other masses vary (figures \ref{fig:set2}a-d) the picture is not significantly affected. This is presented in the example of the large mass splittings $\delta_A,\delta_{\pm}$, however using small $\delta_A$ doesn't make much difference (compare with figure \ref{fig:mz1} noticing the difference in scales).

\paragraph{Low DM mass region}
Since $M_{D_H}$ is small compared to the other masses, $R>1$ is in fact not possible and the possibilities presented in table \ref{raylambda} are reduced to only three rays (Ic, IIa, VI). In the considered range of $M_{D_H}$ the main mass parameter that influences the boundaries of rays is mass of the Higgs, $M_{h_S}$. Changing $M_{D_H}$ in the allowed range with fixed $\delta_A, \delta_\pm$ for this region does not change the picture \ref{low1}. Boundary line between rays IIb and VI depends mostly on $M_{h_S}$ and fixed gauge and Yukawa couplings. Small corrections ($\Delta \lambda_{345} = 0.01$) coming from the change of $\delta_A$ and $ \delta_\pm$ arise for the small ($\lambda_2 \in (0.01,0.1)$) and large ($\lambda_2> 5$) values of $\lambda_2$ .



\begin{figure}[htb]
\centering
\includegraphics[scale=0.4]{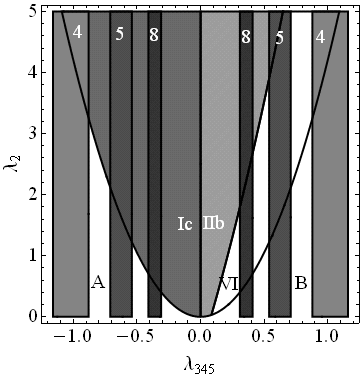}
  \caption{$(\lambda_{345},\lambda_2)$ plot for low DM mass region with the WMAP-allowed areas for $M_{D_H}=(8,5,4)$ GeV, respectively. \label{omega3}}
\end{figure}

In section 3 we presented the $\Omega_{DM} h^2$ calculations for this region. Here we see that the values of $M_{D_H}$ and $\lambda_2$ are the most important parameters.
In a way, $\lambda_2$ is more important than $\lambda_{345}$, because even if we are in the allowed region of $\lambda_{345}$, the small value of $\lambda_2$ may lead us to a region excluded by the positivity constraints. As $M_{D_H}$ grows, we enter the WMAP region for the lower absolute values of $\lambda_{345}$ which corresponds to lower values of $\lambda_2$ (figure \ref{omega3}). 

\paragraph{Medium DM mass region}

For large mass splittings $\delta_A \sim \delta_\pm$  all rays (and all types of evolution) are possible in the range $M_{D_H} = (30-60) \textrm{ GeV}$ (figures \ref{mid11},\ref{fig:s3}-d). As mass grows, ray VI requires higher values of $\lambda_2$ and the region of this ray shrinks in benefit of rays IIa, IV and V. For the medium mass region the change of $M_{D_H}$ effects the evolution for low values of $\lambda_2$ more than the change of $M_{D_A}$ and $M_{D^{\pm}}$ for fixed value of $M_{D_H}$ (figures \ref{fig:set2}a-d).

For small $\delta_A$ the picture is similar to the previous one, however as the $M_{D_H}$ grows the allowed region of ray VI becomes limited in benefit of rays IIa and IV. For $M_{D_H} = 80 \textrm{ GeV}$ ray VI is still possible but it requires higher values of $\lambda_2 \approx 0.7$ (figures \ref{fig:mid12}a-c).

The types of rays that may be realized with the $\lambda_{345}$ inside the WMAP limit depend heavily on the specific value of $M_{D_H}$.

\begin{figure}[h]
\vspace{-10pt}
  \centering
  \subfloat[$M_{D_H} = 65 \textrm{ GeV}$]{\label{fig:mza1}\includegraphics[width=0.23\textwidth]{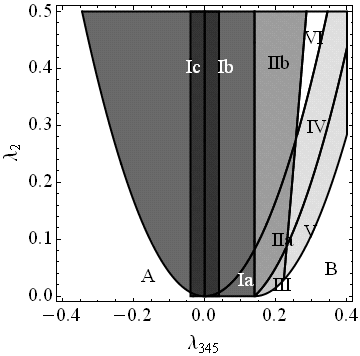}} \;
  \subfloat[$M_{D_H} = 70 \textrm{ GeV}$]{\label{fig:mza2}\includegraphics[width=0.23\textwidth]{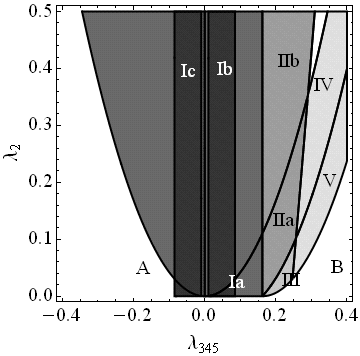}} \;
  \subfloat[$M_{D_H} = 75 \textrm{ GeV}$]{\label{fig:mza3}\includegraphics[width=0.23\textwidth]{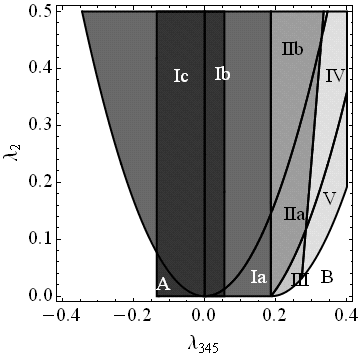}}
\vspace{-5pt}\caption{Medium mass region for $\delta_A = 8 \textrm{ GeV}, \delta_\pm = 50 \textrm{ GeV} $. Vertical dark areas correspond to the relic density in 3$\sigma$ WMAP range.  \label{fig:small2}}
\end{figure}

For small $\delta_A$ region around $\lambda_{345} = 0$ gives proper $\Omega_{DM} h^2$. Here only sequences with a single phase transition can be realized. As $M_{D_H}$ grows, larger $\lambda_{345}$ are allowed by WMAP. However, also the region where the more complex sequences  can be realized shifts towards even larger $\lambda_{345}$ (figure \ref{fig:mid12}). In general, this case may provide us only the $EW\! s \to I_1$ sequences in agreement with WMAP (figure \ref{fig:small2}). 

\begin{figure}[hb]
\vspace{-10pt}
  \centering
  \subfloat[$M_{D_H}=30$ GeV]{\label{fig:za1}\includegraphics[width=0.23\textwidth]{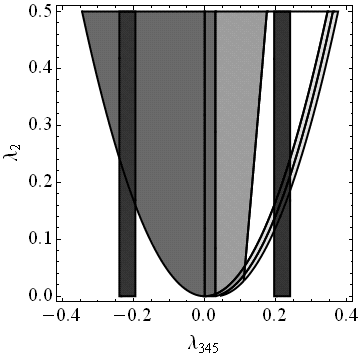}} \;
  \subfloat[$M_{D_H}=35$ GeV]{\label{fig:za2}\includegraphics[width=0.23\textwidth]{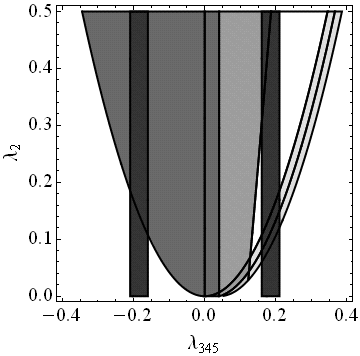}} \;
   \subfloat[$M_{D_H}=40$ GeV]{\label{fig:za3}\includegraphics[width=0.23\textwidth]{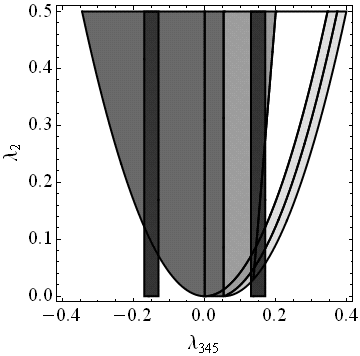}} \;
   \subfloat[$M_{D_H}=50$ GeV]{\label{fig:za4}\includegraphics[width=0.23\textwidth]{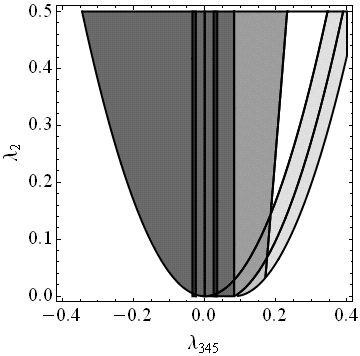}}
\vspace{-5pt}
  \caption{Medium mass region for $\delta_A = 70 \textrm{ GeV}, \delta_\pm = 70 \textrm{ GeV} $. Vertical dark areas correspond to the relic density in 3$\sigma$ WMAP range.  \label{fig:large2}}
\end{figure}

For large $\delta_A$ also sequences III-VI can be realized with the proper relic density (figure \ref{fig:large2}). This is the case for masses $M_{D_H} \approx 40$ GeV, where rays with the 1st-order phase transition are possible. For this value of mass the WMAP allowed range is in $\lambda_{345} \in (0.13,0.17)$. One should keep in mind that this region can be constrained soon by the DM direct detection data. For larger masses the region of those complex sequences is shifted towards the larger values of $\lambda_{345}$. In general they give $\Omega_{DM} h^2$ below the WMAP limit due to the $D_H D_H \to h_S \to \bar{b} b$ channel. In this case we are left only with the $EW\! s \to I_1$ sequences  realized for smaller $\lambda_{345}$ inside WMAP limit.

\paragraph{High DM mass region} In this region the lower value of $M_{D_H}$ makes the realization of other than a single-phase transition types evolution possible for lower values of $\lambda_{345}$. Still, for $M_{D_H} = 500$ GeV this value is high of the order $O(10)$. If we limit ourselves to $\lambda_i \approx O(1)$  then only single phase transition is possible in the high mass region and the $(\lambda_{345},\lambda_2)$ plot is virtually identical to figure \ref{high1}. As discussed before, proper $\Omega_{DM} h^2$ is for $\lambda_{345} < O(1)$ and thus the complex sequences are not in agreement with WMAP.

\section{Conclusions and outlook}

In this work we analysed the possible types of evolution of the Universe (rays) in the IDM, including constraints from the DM relic density data (WMAP). Three regions of the allowed dark matter masses: low, medium and high mass were considered. Evolution wasparametrized by masses of scalars (including dark matter particle) and two parameters (self-couplings) $\lambda_{345}$ and $\lambda_2$. With fixed scalar masses we can find the values of $(\lambda_{345},\lambda_2)$ that satisfy the conditions of a chosen ray (table \ref{raylambda} in the appendix). We have shown that those three regions of DM mass exhibit the different behaviour: both in the possible types of evolution and the energy relic density values.

For the low DM mass region only three kinds of rays can be realized: ray Ic, IIb and VI. This is due to the very low $D_H$ mass and the exact values of the much heavier other scalars do not affect the picture. The energy relic density also doesn't depend on $M_{D_A}$ or $M_{D\pm}$, but it is stronly influenced by the exact value of $M_{D_H}$. In general, rather larger $|\lambda_{345}| \sim (0.4-1)$ are needed to fulfill the WMAP constraints. Due to this fact, in our approximation of the $\Omega_{DM} h^2$ calculation, sequence $EW\! s \to I_2 \to M \to I_1$ gives the proper $\Omega_{DM} h^2$.

For the medium DM mass region all types of evolutions are possible. Again the main parameter that affects the regions in $(\lambda_{345},\lambda_2)$ plane is the mass of DM particle, however for lower $\lambda_2$ also the effect of changing of $M_{D_A},M_{D^\pm}$ are visible. In this region $\Omega_{DM} h^2$ depends heavily on the value of $M_{D_H}$, but also on $\delta_A$. There are two separate regions. For small $\delta_A$ coannihilation is possible and in general small $|\lambda_{345}|$ give the proper DM relic density. Large $\delta_A$ case is similar to the low DM mass case as the coannihilation is not important and the allowed $|\lambda_{345}|$ tends to shift towards the larger values.

In the high DM mass region all dark scalars have almost degenerate and very large masses. This has two main consequences. First, the $(\lambda_{345},\lambda_2)$ plane for $\lambda<O(1)$ is reduced only to the regions corresponding to the simple sequences with a single phase transition. Second, the energy relic density is strongly lowered for most values of $|\lambda_{345}|$, because of the coannihilation.

We would like to comment, that the high mass region requires very careful fixing of the parameters, as it is very sensitive to changing the values of $\delta$'s. This certainly doesn't seem to be natural (\textit{fine-tuning}). From this point of view the most interesting is the medium mass region with the large mass splittings, as here the adjustment of the parameters isn't so strict. 

It is worth to stress the importance of the often neglected $\lambda_2$ self-coupling. We argue, that the astrophysical data should be used to limit the values of this self-coupling. Firstly, the $\Omega_{DM} h^2$ calculation for a fixed $M_{D_H}$ gives the allowed values of the $\lambda_{345}$ parameter. This calculation doesn't depend on the exact value of $\lambda_2$. Secondly, those obained values of $\lambda_{345}$ should be used to constrain the $\lambda_2$ parameter through the positivity constraints or the conditions for the realization of the different rays. Therefore, fixing $\lambda_2$ to an arbitrary value during the calculations may result in the exclusion of the WMAP allowed region. This is especially visible in the low and high DM mass region, where for the $\lambda_{345} <0$ the positivity constraints require  $\lambda_2 > (0.2-0.6)$.

As we mentioned in section 3, the calculations of $\Omega_{DM}h^2$ in this work should be treated only as an estimate. We do not take into account the corrections to those calculations arising from the thermal evolution of the Universe, such as the influence of complex sequences of phase transitions or the effect of the 1-st order phase transitions. Also, in certain ranges of $(\lambda_{345},\lambda_2)$ for low and medium DM mass region the final phase transition into inert vacuum may take place at the lower temperatures, which may affect the freeze-out. Those effects are now under investigation.

\paragraph{Acknowledgments} I would like to thank I. Gizburg, K. Kanishev and M. Krawczyk for cooperation and discussions. I would like to thank M. Krawczyk for reading the manuscript. Work was partly supported by Polish Ministry of Science and Higher Education Grant N N202 230337.

\appendix

\section{Conditions for rays}
Below we present the tables that contain the conditions for the realization of different types of evolution presented in two planes: $(\mu_1,\mu_2)$ in table 2 and $(\lambda_{345},\lambda_2)$ in table 3. Here we use the redefined evolution coefficient: 
\begin{equation}
\tilde c=\tilde c_2/\tilde c_1 =  (c_2 \sqrt{\lambda_1})/(c_1 \sqrt{\lambda_2}).
\end{equation}
To make the table more transparent we use the previously introduced boundary values  $\tilde \lambda_2,\tilde \lambda_{345}$:
\begin{eqnarray}
\tilde \lambda_2 &=&  \frac{-(3 g^2 + g^{\prime 2})}{4}- \frac{ (M_{D_A}^2 - 3 M_{D_H}^2 + 2 M^2_{D^\pm})}{3 v^2} + \frac{\lambda_{345}}{3}  + \frac{(2 M_{D_H}^2 - v^2 \lambda_{345})}{12} \times  \nonumber \\ &&  \times \frac{(-4 M_{D_A}^2 + 12 M_{D_H}^2 - 8 M^2_{D^\pm} - (9 g^2  + 3 g^{\prime 2}  + 12 g_b^2  + 12 g_t^2  + 8  \lambda_{345})v^2)}{v^2 M_{h_s}^2 }.
 \label{l2bondA}
\end{eqnarray}
\begin{equation}
\tilde \lambda_{345} = \frac{2 M_{D_H}^2}{v^2}
\end{equation}
and the abbreviation:
\begin{equation}
\alpha =  \frac{(v^2 \lambda_{345} -2 M_{D_H}^2 )}{M^2_{h_s}}
\end{equation}

\begin{center}
\begin{table}
\centering
{\renewcommand{\arraystretch}{1.5}

\begin{tabular}{|c|c|c|}
\hline
\hline
Ray no. & $R$ & Conditions \\ \hline \hline
\multicolumn{3}{|c|} {$EW\! s \to I_1$}\\ \hline\hline
Ia & $R > 1$ & $\mu_2<0$ \\ \hline
IIa & $R > 1$ & $0 < \mu_2 < \textrm{Min}\left( \mu_1 \tilde c , \mu_1 R^{-1} \right)$ \\ \hline
III & $R > 1$ & $\mu_1 R^{-1} < \mu_2 < \textrm{Min}\left( \mu_1 \tilde c, \mu_1 \right)$  \\ \hline
Ib & $1 > R > 0$ & $\mu_2 <0$ \\ \hline
IIb & $1 > R > 0$ & $0<\mu_2<\textrm{Min}\left( \mu_1 \tilde c, \mu_1 R  \right)$  \\ \hline
Ic & $0 > R > -1$ & $\mu_2 < \mu_1 R < 0$ \\ \hline \hline
\multicolumn{3}{|c|} {$EW\! s \to I_2 \to I_1$}\\ \hline \hline
IV & $R > 1$ & $\mu_1 \tilde c < \mu_2 < \mu_1 R^{-1}$ \\ \hline
V & $R > 1$ & $\textrm{Max}\left( \mu_1 \tilde c, \mu_1 R^{-1} \right) < \mu_2 <\mu_1$ \\ \hline \hline
\multicolumn{3}{|c|} {$EW\! s \to I_2 \to M \to I_1$}\\ \hline \hline
VI & $1 > R > 0$ & $\mu_1  \tilde c <\mu_2<\mu_1 R$ \\ \hline
\hline
\end{tabular}
}
  \caption{Possible rays \label{raym1m2}; ($\mu_1,\mu_2$) conditions.}
\end{table}
\end{center}

\begin{table}
\centering
{\renewcommand{\arraystretch}{1.8}

\begin{tabular}{|c|c|c|}
\hline \hline
Ray no. & $\lambda_{345}$ region & $\lambda_{2}$ region \\ \hline \hline
\multicolumn{3}{|c|} {$EW\! s \to I_1$}\\ \hline\hline
Ia & $0< \lambda_{345}< \tilde \lambda_{345}$ & $\lambda_2 > \frac{\lambda_{345}^2 v^2}{M^2_{h_s}}$ \\ \hline
IIa & $\lambda_{345} > \tilde \lambda_{345}$ & $\textrm{Max} \left( \tilde \lambda_2, \alpha  \lambda_{345} \right) < \lambda_2 < \frac{\lambda_{345}^2 v^2}{M^2_{h_s}}$ \\ \hline
III &  & $\textrm{Max} \left( \tilde \lambda_2, \frac{\alpha^2 M_{h_S}^2}{ v^2} \right) < \lambda_2 < \alpha \lambda_{345}$ \\ \hline
Ib & $0< \lambda_{345}< \tilde \lambda_{345}$ & $\lambda_2 > \frac{\lambda_{345}^2 v^2}{M^2_{h_s}}$ \\ \hline
IIb & $\lambda_{345} > \tilde \lambda_{345}$ & $\textrm{Max} \left( \tilde \lambda_2, \frac{\lambda_{345}^2 v^2}{M^2_{h_s}} \right) < \lambda_2$  \\ \hline
Ic & $\lambda_{345} < 0 $ & $\lambda_2 > \frac{\lambda_{345}^2 v^2}{M^2_{h_s}}$ \\ \hline \hline
\multicolumn{3}{|c|} {$EW\! s \to I_2 \to I_1$}\\ \hline \hline
IV &  & $ \alpha \lambda_{345} < \lambda_2 < \textrm{Min} \left( \tilde \lambda_2, \frac{\lambda_{345}^2 v^2}{M^2_{h_s}} \right) $ \\ \hline
V &  & $\frac{\alpha^2 M_{h_S}^2}{ v^2} < \lambda_2 < \textrm{Min} \left( \tilde \lambda_2, \alpha \lambda_{345} \right)$ \\ \hline \hline
\multicolumn{3}{|c|} {$EW\! s \to I_2 \to M \to I_1$}\\ \hline \hline
VI &  & $\frac{\lambda_{345}^2 v^2}{M^2_{h_s}} < \lambda_2 < \tilde \lambda_2$  \\ \hline
\hline
\end{tabular}
}
  \caption{Possible rays; $(\lambda_{345},\lambda_2)$ conditions. \label{raylambda}}
\end{table}


\begin{thebibliography}{99}

\bibitem{Deshpande:1977rw}
  N.~G.~Deshpande and E.~Ma,
  ``Pattern Of Symmetry Breaking With Two Higgs Doublets,''
  Phys.\ Rev.\  D {\bf 18} (1978) 2574.
\bibitem{Barbieri:2006dq}
  R.~Barbieri, L.~J.~Hall and V.~S.~Rychkov,
  ``Improved naturalness with a heavy Higgs: An alternative road to LHC
  physics,''
  Phys.\ Rev.\  D {\bf 74} (2006) 015007
  [arXiv:hep-ph/0603188].

\bibitem{Ginzburg:2007jn}
  I.~F.~Ginzburg and K.~A.~Kanishev,
  ``Different vacua in 2HDM,''
  Phys.\ Rev.\  D {\bf 76} (2007) 095013
  [arXiv:0704.3664 [hep-ph]].
  
  
\bibitem{Ginzburg:2009dp}
  I.~F.~Ginzburg, I.~P.~Ivanov and K.~A.~Kanishev,
  ``The evolution of vacuum states and phase transitions in 2HDM during cooling
  of Universe,''
  Phys.\ Rev.\  D {\bf 81} (2010) 085031
  [arXiv:0911.2383 [hep-ph]].

\bibitem{Ivanov:2008er}
  I.~P.~Ivanov,
  ``Thermal evolution of the ground state of the most general 2HDM,''
  Acta Phys.\ Polon.\  B {\bf 40} (2009) 2789
  [arXiv:0812.4984 [hep-ph]].

\bibitem{Ginzburg:2010wa}
  I.~F.~Ginzburg, K.~A.~Kanishev, M.~Krawczyk and D.~Sokolowska,
  ``Evolution of Universe to the present inert phase,''
  Phys.\ Rev.\  D {\bf 82}, 123533 (2010)
  [arXiv:1009.4593 [hep-ph]].

\bibitem{nasza_druga} D. Sokolowska, ``Temperature evolution of physical parameters in the Inert Doublet Model,'' arXiv:1104.3326 [hep-ph]




\bibitem{Cao:2007rm}
  Q.~H.~Cao, E.~Ma and G.~Rajasekaran,
  ``Observing the Dark Scalar Doublet and its Impact on the Standard-Model
  Higgs Boson at Colliders,''
  Phys.\ Rev.\  D {\bf 76} (2007) 095011
  [arXiv:0708.2939 [hep-ph]].
\bibitem{Agrawal:2008xz}
  P.~Agrawal, E.~M.~Dolle and C.~A.~Krenke,
  ``Signals of Inert Doublet Dark Matter in Neutrino Telescopes,''
  Phys.\ Rev.\  D {\bf 79}, 015015 (2009)
  [arXiv:0811.1798 [hep-ph]].
\bibitem{Gustafsson:2007pc}
  M.~Gustafsson, E.~Lundstrom, L.~Bergstrom and J.~Edsjo,
  ``Significant gamma lines from inert Higgs dark matter,''
  Phys.\ Rev.\ Lett.\  {\bf 99} (2007) 041301
  [arXiv:astro-ph/0703512].


\bibitem{Dolle:2009fn}
  E.~M.~Dolle and S.~Su,
  ``The Inert Dark Matter,''
  Phys.\ Rev.\  D {\bf 80} (2009) 055012
  [arXiv:0906.1609 [hep-ph]].
  
\bibitem{Dolle:2009ft}
  E.~Dolle, X.~Miao, S.~Su and B.~Thomas,
  ``Dilepton Signals in the Inert Doublet Model,''
  Phys.\ Rev.\  D {\bf 81}, 035003 (2010)
  [arXiv:0909.3094 [hep-ph]].

\bibitem{LopezHonorez:2006gr}
  L.~Lopez Honorez, E.~Nezri, J.~F.~Oliver and M.~H.~G.~Tytgat,
  ``The inert doublet model: An archetype for dark matter,''
  JCAP {\bf 0702} (2007) 028
  [arXiv:hep-ph/0612275].

\bibitem{Arina:2009um}
  C.~Arina, F.~S.~Ling and M.~H.~G.~Tytgat,
  ``IDM and iDM or The Inert Doublet Model and Inelastic Dark Matter,''
  JCAP {\bf 0910}, 018 (2009)
  [arXiv:0907.0430 [hep-ph]].

\bibitem{Tytgat:2007cv}
  M.~H.~G.~Tytgat,
  ``The Inert Doublet Model : a new archetype of WIMP dark matter?,''
  J.\ Phys.\ Conf.\ Ser.\  {\bf 120} (2008) 042026
  [arXiv:0712.4206 [hep-ph]].
\bibitem{Honorez:2010re}
  L.~L.~Honorez and C.~E.~Yaguna,
  ``The inert doublet model of dark matter revisited,''
  arXiv:1003.3125 [hep-ph].

\bibitem{Lundstrom:2008ai}
  E.~Lundstrom, M.~Gustafsson and J.~Edsjo,
  ``The Inert Doublet Model and LEP II Limits,''
  Phys.\ Rev.\  D {\bf 79} (2009) 035013
  [arXiv:0810.3924 [hep-ph]].
  
  
\bibitem{Krawczyk:2009fb}
  M.~Krawczyk and D.~Soko\l owska,
  ``Constraining the Dark 2HDM,''
  arXiv:0911.2457 [hep-ph].

\bibitem{PDG} Particle Data Group. {\it Journ. of Phys.} {\bf G 37} \#7A (2010) 075021.

\bibitem{Belanger:2010gh}
  G.~Belanger, F.~Boudjema, P.~Brun, A.~Pukhov, S.~Rosier-Lees, P.~Salati and A.~Semenov,
  ``Indirect search for dark matter with micrOMEGAs2.4,''
  Comput.\ Phys.\ Commun.\  {\bf 182}, 842 (2011)
  [arXiv:1004.1092 [hep-ph]].


\bibitem{pracaTEMP} I. Ginzburg (private communication); I.Ginzburg, K.Kanishev, M.Krawczyk, D.Soko\l owska (work in progress).

\bibitem{Gavela:1998ux}
  M.~B.~Gavela, O.~Pene, N.~Rius and S.~Vargas-Castrillon,
  ``The fading of symmetry non-restoration at finite temperature,''
  Phys.\ Rev.\  D {\bf 59} (1999) 025008
  [arXiv:hep-ph/9801244].





%
%
%
%
%
%
%
%
\end{thebibliography}
\end{document}